\documentclass[sigconf]{acmart}

\usepackage{rotating}
\usepackage{array}
\usepackage{url}
\usepackage{xspace}
\usepackage{pifont}
\usepackage{adjustbox}
\usepackage{booktabs, multirow}
\usepackage{soul}
\usepackage{xcolor,colortbl}
\usepackage{changepage,threeparttable}
\usepackage{graphicx}
\usepackage{etoolbox}
\usepackage{pgf}\usepackage{enumitem}
\usepackage{amsmath}
\usepackage{makecell}
\usepackage{tcolorbox}
\usepackage{comment}
\usepackage{pgfplotstable}
\usepackage{tikz}
\usepackage{tabularx}
\usepackage{pdfpages}
\usepackage{esvect}
\usepackage{lscape}
\usepackage{mdframed}
\usepackage{amsthm}
\usepackage{caption}

\usepackage{blindtext}

\usepackage{etoolbox}
\makeatletter
\patchcmd{\maketitle}{\@copyrightspace}{}{}{}
\makeatother

\newcommand{\BfPara}[1]{\noindent \textbf{#1.}\xspace}
\newcommand{\xl}[1]{\Xhline{#1\arrayrulewidth}}
\newcommand{\etal}{{\em et al.}\xspace}
\newcommand{\eg}[1]{{e.g.,}\xspace}
\newcommand{\ie}[1]{{i.e.,}\xspace}

\newcommand{\cmark}{\textcolor{green!70!black}{\ding{51}}}

\definecolor{boxcolor}{RGB}{240, 248, 255} 

\newmdenv[
  backgroundcolor=boxcolor,
  linewidth=0.25pt,
  innerleftmargin=3pt,
  innerrightmargin=3pt,
  innertopmargin=1pt,
  innerbottommargin=1pt
]{takeawaybox}

\setcopyright{none}
\renewcommand\footnotetextcopyrightpermission[1]{} 
\newtheorem{takeaway}{Takeaway}

\title{Large Language Models in Cybersecurity: Applications, Vulnerabilities, and Defense Techniques}

\author{Niveen O. Jaffal}
\affiliation{
  \institution{Birzeit University}
  \country{Palestine}
}
\email{njaffal@birzeit.edu}

\author{Mohammed Alkhanafseh}
\affiliation{
  \institution{Birzeit University}
  \country{Palestine}
}
\email{mkhanafseh@@birzeit.edu}

\author{David Mohaisen}
\authornote{Corresponding author}
\affiliation{
  \institution{University of Central Florida}
  \country{USA}
}
\email{mohaisen@ucf.edu}

\begin{document}

\begin{abstract}
Large Language Models (LLMs) are transforming cybersecurity by enabling intelligent, adaptive, and automated approaches to threat detection, vulnerability assessment, and incident response. With their advanced language understanding and contextual reasoning, LLMs surpass traditional methods in tackling challenges across domains such as IoT, blockchain, and hardware security. This survey provides a comprehensive overview of LLM applications in cybersecurity, focusing on two core areas: (1) the integration of LLMs into key cybersecurity domains, and (2) the vulnerabilities of LLMs themselves, along with mitigation strategies. By synthesizing recent advancements and identifying key limitations, this work offers practical insights and strategic recommendations for leveraging LLMs to build secure, scalable, and future-ready cyber defense systems.
\end{abstract}


\maketitle

\section{Introduction}

Advances in machine learning and deep learning, particularly with the advent of transformer architectures \cite{vaswani2023}, have driven the development of LLMs. These advanced Natural Language Processing (NLP) systems, characterized by their extensive parameterization, are trained on foundational tasks such as masked language modeling and autoregressive prediction. These training paradigms enable LLMs to process human language effectively, analyzing contextual semantics and probabilistic relationships across massive text datasets. LLMs exhibit four essential characteristics: (i) a deep understanding of the natural language context; (ii) the ability to generate human-like text; (iii) advanced contextual awareness, particularly in knowledge-intensive applications; and (iv) strong instruction following capabilities that support problem solving and decision making. Prominent LLMs, including BERT \cite{devlin2019}, GPT-3.5~\cite{openai2022gpt35} and GPT-4 ~\cite{openai2023gpt4}, PaLM \cite{chen2024}, Claude~\cite{bai2022}, and Chinchilla \cite{houssel2024}, have demonstrated exceptional performance in various NLP tasks, such as language understanding, text generation, and reasoning. The adaptability of these models is particularly notable as they enable breakthroughs in downstream applications with minimal fine-tuning. These include open domain question answering~\cite{abdallah2024}, dialogue systems~\cite{ou2024}, and program synthesis~\cite{aguinakang2024}, among others. Using rich linguistic representations and robust reasoning capabilities, LLMs are reshaping how NLP challenges are addressed, paving the way for transformative advancements across various domains.

The increasing complexity and sophistication of cyber threats require innovative and adaptive approaches to strengthen cybersecurity mitigation~\cite{DBLP:conf/www/MohaisenA13,DBLP:journals/compsec/MohaisenAM15,DBLP:journals/compsec/JangKWMK16,DBLP:conf/icdcs/ShenVMKZ17,DBLP:conf/edge/ChoiAASNM19,DBLP:conf/icdcs/AbusnainaKA0AM19,DBLP:journals/iotj/AlasmaryKAPCAAN19,DBLP:journals/tmc/ShenVMKZ19,DBLP:conf/icdcs/AlasmaryAJAANM20,DBLP:conf/icics/AnwarAPWCM20,DBLP:conf/dsn/AbusnainaAAAJNM21,DBLP:conf/raid/AbusnainaAAAJNM22,DBLP:journals/cn/ChoiAAASWCNAM22,DBLP:journals/tdsc/AbusnainaAAAJSN22,DBLP:conf/wise/AbusnainaASAJSM24,DBLP:journals/tdsc/ZapzalkaSM25,AlghamdiM24}, and LLMs have found many applications in the security domain~\cite{KharmaCAM25}. For instance, LLMs are known to generalize from human languages to other domains with minor modifications, making them ideal for automating the generation of security rules, associating cyber threats with one another, and even discovering new phenomena and threats unseen before. Moreover, the fact that such models are deployed in a rather hostile environment with both begin and malicious users makes them ideal for manipulation, giving rise to the idea of understanding the robustness of such models through a finer understanding of their attack surface and mitigations to minimize such surface. This survey offers a comprehensive analysis of the transformative impact that LLMs could have on the field of cybersecurity by identifying the various applications of such tools, enumerating their vulnerabilities, and highlighting key defense techniques. In compiling this survey, we hope to shed light on the existing literature and identify gaps in such literature that could open the door for future work. 

Although there have been several surveys in this space, ours stands out in multiple ways. First, it is the first survey to integrate the application landscape in cybersecurity with the attack surface and defense mechanisms, providing a comprehensive enumeration of a broader spectrum of applications. Second, given the evolving nature of this domain, our survey encompasses a range of studies across previously unexplored dimensions.

\subsection{Research Questions} To effectively position our work in the context of previous works, we formulate several broad research questions that we aim to address through this effort.  {\bf RQ 1. What are the key cybersecurity domains where LLMs address specific tasks and challenges within these domains and how?} The first research question focuses on the scope and nature of security tasks in which LLMs have been applied, with the aim of categorizing and understanding the breadth of security challenges addressed in different security domains. By analyzing previous studies, this question seeks to provide a detailed inventory of the various security tasks that utilize LLM, offering insight into their adaptability, effectiveness, and impact within each domain. {\bf RQ 2. What vulnerabilities are associated with LLM in cyber security applications and what strategies can be implemented to mitigate these risks and protect models?} The second research question investigates the vulnerabilities of LLMs and explores defense techniques to improve their security. By analyzing potential attack vectors and prevention techniques, it provides a comprehensive understanding of the challenges and safeguards necessary to improve the resilience of these models in cybersecurity applications.

\subsection{Contributions} The primary objective of this survey is to explore the future of cybersecurity through the lens of generative Artificial Intelligence (AI) and LLMs, encompassing all the main critical aspects of the cyber domain. To this end, the key contributions of this survey are summarized as follows:

\begin{itemize}
\item {\bf Applications.} We enumerate and evaluate the diverse applications of LLMs in cybersecurity, including hardware design security, intrusion detection, malware detection, and phishing prevention, while analyzing their capabilities in these different contexts to address how LLMs are used.

\item {\bf Vulnerabilities and Mitigation.} We systematically enumerate and examine the vulnerabilities in LLMs with respect to their implications for security applications. Our exploration of such an attack surface encompasses aspects like prompt injection, jailbreaking attack, data poisoning, and backdoor attacks. Moreover, we enumerate and assess the defense techniques that are in place or could be further deployed to reduce these risks and improve LLMs security for those critical applications.
\item {\bf Potential Challenges.} We identify the possible potential challenges that arise in the use of LLMs for specified cybersecurity tasks, calling for more attention and research actions from the community.
\end{itemize}

\subsection{Organization} 

The remaining sections are structured as follows. Section~\ref{sec:related} summarizes recent surveys on LLMs and their applications in various cybersecurity domains, highlighting previous work and positioning this survey within the literature on LLM applications and security. Section~\ref{sec:Security_Domains_and_Tasks} explores the security domains enhanced by LLM innovations, analyzing their impact on cybersecurity applications and their effectiveness in addressing complex challenges. Section~\ref{sec:Vulnerabilities_and_Mitigation_Strategies} investigates LLM vulnerabilities, categorizing threats and countermeasures while highlighting the need for proactive mitigation strategies to ensure secure deployment. Section~\ref{sec:Limitations} discusses the challenges and limitations of integrating LLMs into cybersecurity, considering both practical and theoretical aspects. Finally, Section~\ref{sec:Conclusion} highlights the potential of LLMs in cybersecurity, summarizes key findings and defense strategies, and outlines research directions.

\section{Methodology}

\begin{figure}[t]
    \centering
    \includegraphics[width=.99\linewidth]{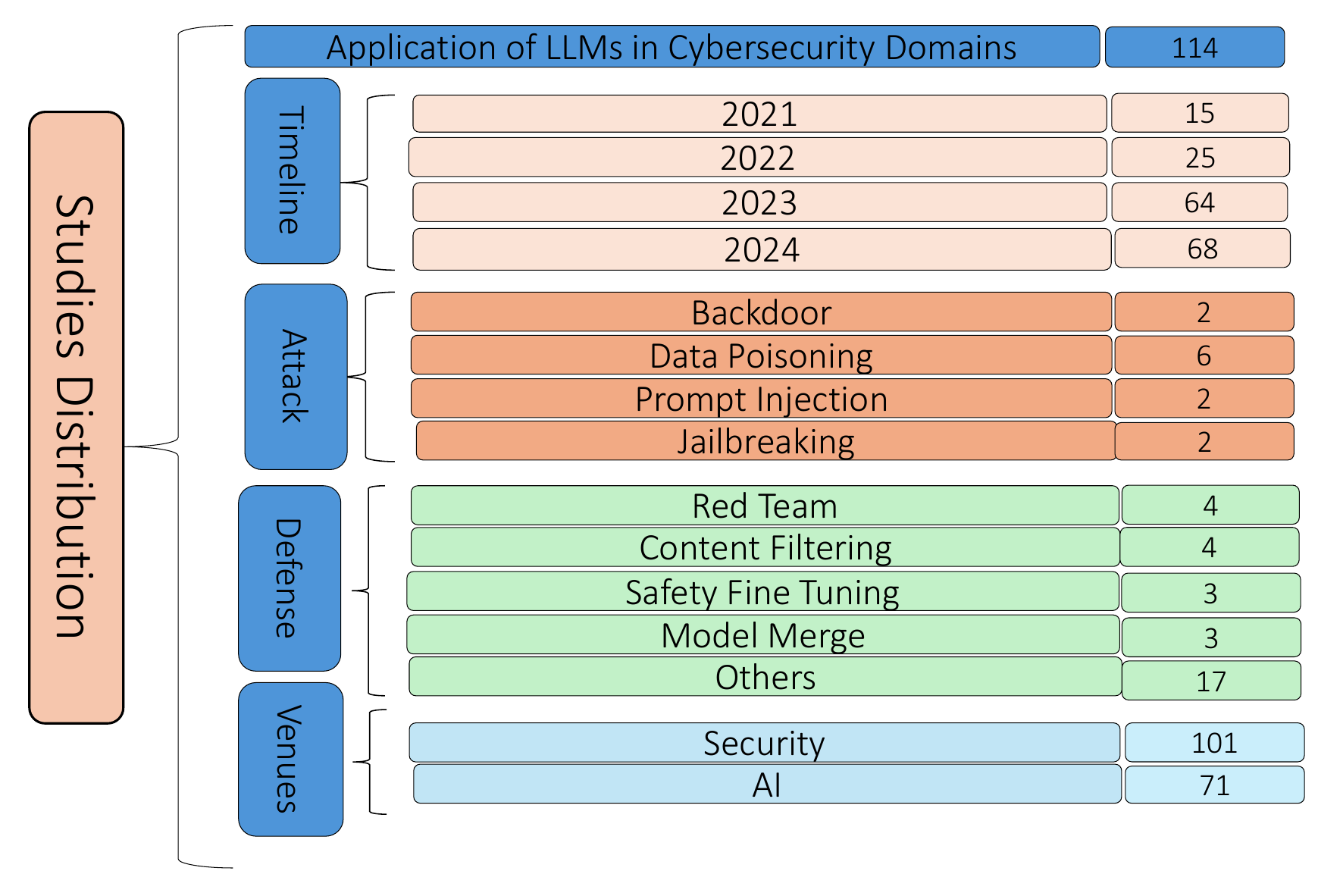}
    \caption{Comprehensive Survey Studies Analysis}
    \label{Figure}
\end{figure}

To conduct this survey, we prioritize key factors in selecting relevant studies, including timeliness, coverage, diversity, and relevance. For timeliness, we focus on research published within the last four years, emphasizing the most recent three. This selection includes 15 studies from 2021, 25 from 2022, 64 from 2023, and 68 from 2024. For coverage, we ensure a broad range of applications, attack vectors, and defense mechanisms, encompassing 114 studies on applications, 12 on attacks, and 31 on defense techniques. To enhance diversity, we include studies from both AI (71 studies) and security research venues (101 studies). Figure~\ref{Figure} visualizes this distribution, illustrating the evolution of research over time, the dominance of applications, attacks, and defenses in selected studies, and the representation of AI and security venues. This structured approach ensures a comprehensive analysis.

\section{Related Reviews}\label{sec:related} We present in this section several previously selected surveys that make significant contributions; these surveys encompass a detailed and thorough examination of key aspects related to LLM, including the datasets utilized for training and fine-tuning these models, the vulnerabilities inherent to these models, and the strategies proposed for their mitigations. In addition, they highlight the innovative methodologies employed by LLMs to address complex security challenges, offering a comparative analysis of their efficacy in addressing issues across domains such as cloud computing, the Internet of Things (IoT),  hardware, Blockchain, software, and system security.

\begin{table*}[t]
\caption{
A summary of the related work of LLMs applications. Highlights study contribution, \underline{D}atasets used by LLM for security use case, \underline{V}ulnerabilities associated with LLMs and defense techniques, \underline{O}ptimized techniques for LLMs, and domains: \ding{172} Internet of Things, \ding{173} Cloud, \ding{174} Hardware, \ding{175} Blockchain, \ding{176} Software and System Security.
}\vspace{-3mm}
\scalebox{0.9}{
\begin{tabular}{llp{3cm}p{9cm}llllllll}
\xl{2}
\multirow{2}{*}{\textbf{Ref}} &
\multirow{2}{*}{\textbf{Year}} &
\multirow{2}{*}{\textbf{Study Area}} &
\multirow{2}{*}{\textbf{Contribution}} &
\multirow{2}{*}{\textbf{D}} &
\multirow{2}{*}{\textbf{V}} &
\multirow{2}{*}{\textbf{O}} &
\multicolumn{5}{c}{\textbf{Domains}}  \\
\cline{8-12}
&  & & & & & & \ding{172} & \ding{173} & \ding{174} & \ding{175} & \ding{176} \\
 \xl{1}
\cite{saha2023}&2023&LLMs in SoC&Surveys LLM potential, challenges, and outlook for SoC security. &\ding{53} &\cmark &\cmark &\ding{53} &\ding{53} &\cmark &\ding{53} &\cmark\\ \hline

\cite{he2024}&2024&LLMs for blockchain &Covers LLM use in auditing, anomaly detection, and vulnerability repair. &\ding{53} &\cmark &\cmark &\ding{53} &\ding{53} &\ding{53} &\cmark &\ding{53}\\ 
\hline
\cite{xu2024a}&2024&LLM applications &Explores use cases, dataset limitations, and mitigation strategies. &\cmark &\ding{53} &\cmark &\ding{53} &\ding{53} &\ding{53} &\ding{53} &\cmark\\ 
\hline
\cite{yigit2024}&2024&GenAI in cybersecurity &Reviews GenAI attack applications in cybersecurity. &\cmark &\cmark &\ding{53} &\ding{53} &\ding{53} &\ding{53} &\ding{53} &\cmark\\ 
\hline
\cite{divakaran2024} &2024 &LLM in cybersecurity &Highlights LLM capabilities for solving key cybersecurity issues. &\cmark &\cmark &\ding{53} &\cmark &\ding{53} &\ding{53} &\ding{53} &\ding{53}\\ 
\hline
\cite{ferrag2024} &2024 &Survey &Surveys 42 models, their roles in cybersecurity, and known flaws. &\cmark &\cmark &\cmark &\ding{53} &\cmark &\cmark &\ding{53} &\cmark\\ 
\xl{1}
Ours&2025 &Apps, vulns., and defenses &Presents LLM use across domains, identifies vulnerabilities, and proposes defenses. &\cmark &\cmark &\cmark &\cmark &\cmark &\cmark &\cmark &\cmark\\ 
\xl{2}
\end{tabular}}
\label{tab:summary}\vspace{-3mm}
\end{table*}

As shown in~\autoref{tab:summary}, this work provides a comprehensive analysis of LLM applications in cybersecurity, distinguishing itself from previous work by addressing critical gaps across five key security domains. Although existing surveys, such as \cite{xu2024a} and \cite{yigit2024}, explore general LLM applications and vulnerabilities, they often lack detailed insights into specific domains such as IoT, cloud, hardware, and blockchain security. Furthermore, our bridges gaps in underexplored domains such as IoT and cloud environments, providing a more holistic view of LLM deployment in cybersecurity. This work advances the state of the art by integrating broader perspectives with practical insights, positioning it as a critical addition to the evolving body of research. Finally, the work covers a range of recent advances and papers that are more timely and were not covered in previous studies and surveys. 

\section{Security Domains and Tasks Empowered by LLM-Based Innovations}
\label{sec:Security_Domains_and_Tasks}
\begin{figure}[t]
    \centering
    \includegraphics[width=.8\linewidth]{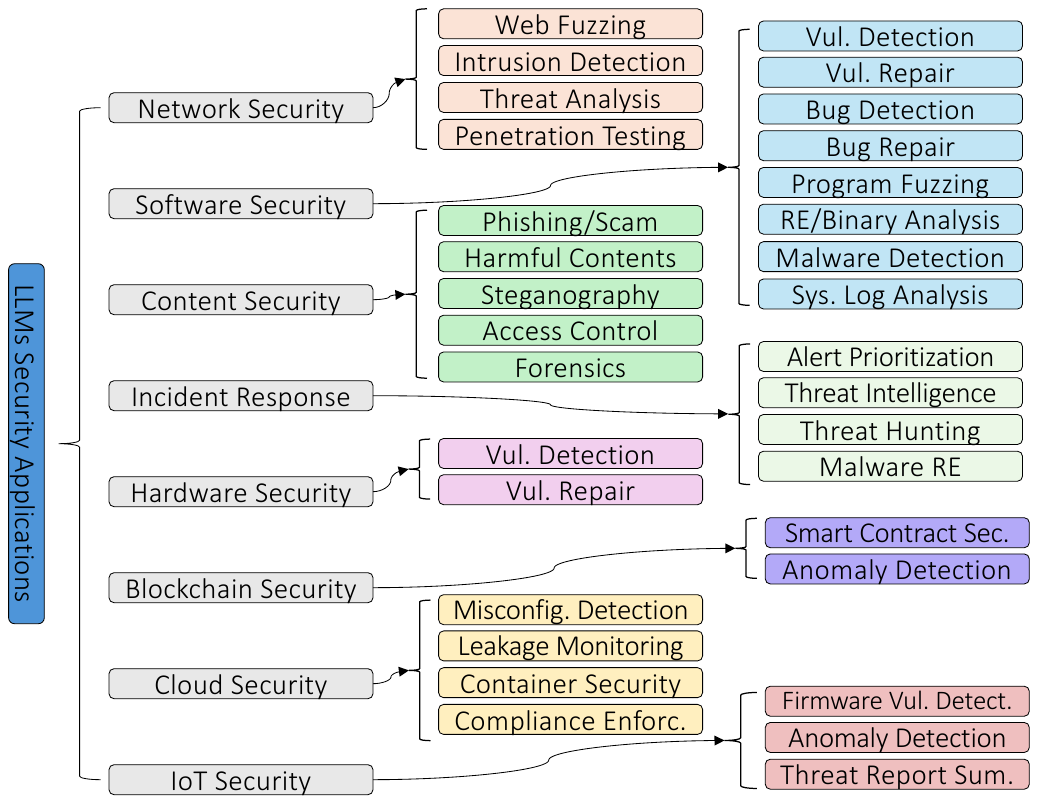}
    \caption{Cybersecurity Domains and LLM Based Applications}
    \label{Figure One}
\end{figure}

The increasing complexity of cybersecurity threats has driven the adoption of LLMs across various security domains. This survey categorizes LLM applications into eight key domains: network, software and system, information and content, hardware, blockchain, cloud, incident response and threat intelligence, and IoT security. This classification provides a structural perspective on the diverse applications of LLMs in cybersecurity domains, offering deeper insights into their contributions to strength security robustness across real-time applications and domains. \autoref{Figure One} visualizes the distribution of LLM applications across cybersecurity domains. Moreover, Tables~\ref{tab:related_work2new} and \ref{tab:related_worknew} present LLM-based solutions for various cybersecurity tasks and use cases. LLMs have proven to be instrumental in improving efficiency, accuracy, and adaptability, marking a significant leap in modern cybersecurity defense strategies.

\subsection{LLMs in Network Security}
This section delves into the applications of LLMs in the field of network security. LLMs have proven to be versatile and powerful tools for addressing a wide range of tasks in this domain, including web fuzzing, traffic and intrusion detection, threat analysis, and penetration testing, which we review in the following subsections. 

\begin{takeaway}
 LLMs enhance network security by improving intrusion and anomaly detection through in-context learning and graph-based techniques. In CTI, they automate intelligence extraction for real-time threat monitoring. Tools like GPTFuzzer refine web fuzzing by generating targeted test cases, while PentestGPT streamlines penetration testing through automated reconnaissance and exploit generation. These advancements boost efficiency, accuracy, and adaptability in network security operations.\end{takeaway}

\subsubsection{Web Fuzzing} Web fuzzing is a mutation-based testing technique that incrementally generates test cases by leveraging coverage feedback obtained from instrumented web applications. Given the critical importance of security in web applications, fuzzing is vital in identifying potential vulnerabilities. For example, Liang~\etal \cite{Liang} introduced GPTFuzzer, a tool built on an encoder-decoder architecture. GPTFuzzer effectively generates payloads targeting Web Application Firewalls (WAFs) to identify and test Structured Query Language injection (SQLi), Cross-Site Scripting (XSS), and Remote Code Execution (RCE) attacks. This is achieved through reinforcement learning, fine-tuning, and applying a KL divergence penalty, which helps overcome local optima during the payload generation process.

Similarly, Liu~\etal\cite{Liu2020} utilized an encoder-decoder architecture to design SQL injection detection test cases for web applications, translating user input into new test scenarios. Meng~\etal\cite{MengMBR24} expands the scope by employing LLMs to generate structured and sequential test inputs for network protocols that lack machine-readable formats. These advancements demonstrate the growing potential of machine learning and fuzzing in enhancing the security testing process for web applications.

\begin{table*}[h!]
\setlength{\tabcolsep}{3pt}
\caption{LLMs-based Models for Cybersecurity (Part I)}
\scalebox{0.8}{
\begin{tabular}{lllp{3cm}p{3cm}p{4.5cm}p{4.2cm}p{4.2cm}}
\xl{2}
\textbf{Author} & \textbf{Ref.} & \textbf{Year} & \textbf{Dataset} & \textbf{Task} & \textbf{Key Contributions} & \textbf{Challenges} & \textbf{Optimized Technique} \\
\xl{1}
Meng~\etal & \cite{MengMBR24} & 2024 & PROFUZZ & Web Fuzzing & Enhanced state transitions and vuln. discovery. & Weak for closed protocols. & GPT-3.5 for grammar and seed tuning. \\
\hline
Liu~\etal & \cite{lanka2024} & 2023 & GramBeddings, Mendeley & Traffic Detection & CharBERT improves URL detection. & High compute, few adversarial tests. & CharBERT + attention + pooling. \\
\hline
Moskal~\etal & \cite{MoskalLHO23} & 2023 & Sim. forensic exp. & Threat Analysis & Refined ACT loop via sandboxed LLM agents. & Weakness with complex net/env. & FSM + chained prompts. \\
\hline
Temara~\etal & \cite{Temara23} & 2023 & None & Pentesting & Multi-tool consolidation via LLM. & Pre-2021 limit, prompt sensitivity. & Case-driven extraction. \\
\hline
Tihanyi~\etal & \cite{TihanyiBJFCM23} & 2023 & Form AI & Vuln. Detection & Benchmarked form AI dataset. & Limited scope, costly verification. & Zero-shot + ESBMC. \\
\hline
Sun~\etal & \cite{SunWXLW24} & 2024 & Solidity, Java CWE & Firmware Vuln. & LLM4Vuln detects misuse + zero-days. & No cross-language support. & Prompt + retrieval + reasoning. \\
\hline
Meng~\etal & \cite{MengSARHP23} & 2023 & RISC, MIPS & HW Vuln. & NSPG w/ HS-BERT for spec extraction. & Doc scarcity, manual labels. & Fine-tuned HS-BERT + MLM. \\
\hline
Du~\etal & \cite{DuY23} & 2023 & Synthetic SFG & Bug Localization & Graph-based CodeBERT + contrastive loss. & Low scale, limited eval. & HMCBL w/ neg. sampling. \\
\hline
Joyce~\etal & \cite{JoycePNR23} & 2023 & VirusTotal (2006-23) & Malware Feature Learn. & AVScan2Vec for AV task embedding. & Mix of benign/malware, retrain needs. & Token + label masking w/ Siamese fine-tuning. \\
\hline
Labonne~\etal & \cite{LabonneM23} & 2023 & Email spam datasets & Phishing Detect. & Spam-T5 outperforms baseline. & Costly training, weak domain generality. & Prefix-tuned Flan-T5. \\
\hline
Malul~\etal & \cite{MalulMMES24} & 2024 & KCFs & Misconfig. Detection & GenKubeSec w/ UMI for high accuracy. & Unseen configs, labeled data, scaling. & Fine-tuned LLM + few-shot. \\
\hline
Kasula~\etal & \cite{kasula2024} & 2023 & NSL-KDD, cloud logs & Data Leakage & Real-time detection in dynamic clouds. & Low generalization. & RF + LSTM. \\
\hline
Bandara~\etal & \cite{Bandara2024} & 2024 & PBOMs & Container Sec. & DevSec-GPT tracks vulns via blockchain. & LLM overhead. & Llama2 + JSON schema. \\
\hline
Nguyen~\etal & \cite{ollabench2024} & 2024 & Compliance data & Compliance & Olla Bench tests LLM reasoning. & Scalability, adaptability. & KG + SEM. \\
\hline
Ji~\etal & \cite{ji2024} & 2024 & Incident reports & Alert Prioritization & SEVENLLM for multi-task response. & Limited multilinguality. & Prompt-tuned LLM. \\
\hline
Gai~\etal & \cite{LuoLV24} & 2023 & Ethereum txns & Anomaly Detect. & BLOCKGPT ranks real-time anomalies. & False positives. & EVM tree + transformer. \\
\hline
Ahmad~\etal & \cite{AhmadTTKP23} & 2023 & MITRE, OpenTitan & HW Bug Repair & LLM repair in Verilog, beats CirFix. & Multi-line bug limits. & Prompt tuning + testbench. \\
\hline
Tseng~\etal & \cite{tseng2024} & 2024 & CTI Reports & Threat Intel & GPT-4 extracts IoCs + Regex rules. & Extraction accuracy. & Segmentation + GPT-4. \\
\hline
Scanlon~\etal & \cite{ScanlonBHHS23} & 2023 & Forensics & Forensic Tasks & GPT-4 used in education + analysis. & Output inconsistency. & Prompt + expert check. \\
\hline
Martin~\etal & \cite{Martin23} & 2023 & OpenAI corpus & Bias Detect. & ChatGPT shows political bias. & Ethical balance. & GSS-based bias reverse. \\
\xl{2}
\end{tabular}}
\label{tab:related_worknew}\label{tab:related_work2new}
\end{table*}

\subsubsection{Traffic and Intrusion detection}

Detecting network traffic and intrusions is crucial for network security and management. LLMs have become powerful tools for intrusion detection, demonstrating versatility across traditional web applications, IoT ecosystems, and in-vehicle networks. These models effectively learn complex patterns in malicious traffic, identify deviations in user behavior that signal anomalies, and interpret the intent behind intrusions and abnormal activities.

A marked improvement is the effort of Liu~\etal\cite{LiuWXQLC23}, who used LLM as a tool to extract features from malicious URLs while expanding the detection to user-level contexts. Zhang~\etal\cite{zhang2024a} demonstrated that employing GPT-4 for in-context learning can achieve an intrusion detection accuracy of more than 95\% with a limited amount of labeled data, thus eliminating the need for fine-tuning. Houssel~\etal\cite{houssel2024} explored the explainability of LLMs in the context of NetFlow-based Network Intrusion Detection Systems (NIDS), showing that they can augment traditional methods and improve interpretability by using tools such as Retrieval Augmented Generation (RAG). LLMs promise real-time analysis and response, enabling zero-day vulnerability discovery and identification of emerging attack patterns through continuous learning. Additionally, they scale efficiently to handle large network traffic, making them ideal for high throughput settings. Their integration with methods such as graph-based anomaly detection and reinforcement learning increases their ability to detect complex threats. These capabilities highlight the crucial role of LLMs in modern intrusion detection, providing robust, scalable, and advanced digital infrastructure protection.

\subsubsection{Cyber Threat Intelligence (CTI)} CTI
is now a vital element in risk management, as highlighted by recent studies \cite{Chen}. The rise in CTI reports requires automated tools for efficient creation and assessment. In network threat analysis, LLMs are critical, particularly in CTI generation and analysis, enhancing decision making processes. The CTI generation involves extracting intelligence from diverse sources, such as books, blogs, and news, and transforming it into structured reports. An example is CVEDrill by Aghaei~\etal~\cite{aghaei2023}, which helps to formulate prioritized cybersecurity risk reports and predict their influence on systems. Similarly, Moskal~\etal\cite{MoskalLHO23} explored the role of ChatGPT in automating responses to network threats, demonstrating its utility in handling basic attack scenarios.

LLMs improve CTI reports by providing real-time updates to monitor evolving cyber threats. Their ability to handle large datasets, identify threat patterns, and merge intelligence from multiple sources improves the efficiency and effectiveness of cybersecurity efforts. With capabilities such as integrating live threat data, modeling attack scenarios, and offering predictive insights, LLMs play a vital role in proactive threat management, advancing CTI reporting, and supporting informed, data-driven decisions in cybersecurity.

\subsubsection{Penetration Testing} Penetration testing is identified by simulated attacks on computer systems to assess their defenses and remains essential for organizations in combating cyber threats. Traditionally, penetration testing involves three key stages: information gathering, payload creation, and vulnerability exploitation. Recent advances emphasize the crucial role of LLMs in automating and improving these steps. Temara~\etal\cite{Temara23} utilized LLMs to optimize data collection by retrieving key information about the target, such as IP addresses, domain info, vendor technologies, and Secure Socket Layer (SSL) / Transport Layer Security (TLS) certificates. Likewise, Sai Charan~\etal \cite{charan2023} explored LLMs in the creation of malicious payloads, noting that models such as ChatGPT can produce more precise payloads, underscoring the challenge of dual use of the technology. Additionally, Happe~\etal\cite{happe2024} advanced the Linux privilege escalation automation with LLM, with practical guidance for privilege escalation in penetration tests.

PentestGPT \cite{deng2024} is a cutting-edge automated penetration testing tool that uses LLMs, demonstrating remarkable abilities in a benchmark with 13 scenarios and 182 subtasks. Its effectiveness comes from three self-interacting modules: inference, generation, and parsing, which enhance task management and response to complex tests. Investigations are underway into the use of LLMs for adversarial simulations and custom exploit scripts, valuable for practical testing. Future developments could include reinforcement learning and adaptive features in real-time to enhance the simulation of dynamic threats and improve the accuracy and prediction of testing. However, the risk of misuse underscores the need for responsible use and strict ethical oversight.

\begin{takeaway}
\BfPara{Challenges and Open Directions} Despite their advantages, LLMs for network security face key challenges, including vulnerability to adversarial attacks, raising concerns about robustness and security. Real-time adaptability requires efficient continuous learning while maintaining scalability in high-throughput environments. Ethical risks, particularly dual use concerns, require safeguards against malicious exploitation. In addition, enhancing explainability and transparency in security-critical applications is vital to trust and informed decision making. Addressing these issues is crucial to fully harnessing LLMs for network security.\end{takeaway}

\subsection {LLMs in Software and System Security}
The complexity in software systems coupled with the increasing number of reported vulnerabilities require advanced security solutions. LLMs have emerged as powerful tools for automating vulnerability, malware detection, program fuzzing, bug repair, reverse engineering, binary analysis, and system log analysis. These tasks are important for ensuring software reliability, improving automated security assessments, and mitigating security threats. Furthermore, their ability for processing and analyzing a large amount of code and large-scale system logs outperform traditional approaches, specially in enabling real-time anomaly detection and proactive threat mitigation. In this section, we explore and examine how LLMs improve software and security tasks by introducing advanced techniques that improve accuracy, efficiency, and scalability.
\begin{takeaway}
 LLMs enhance software and system security by enabling real-time vulnerability detection, automated bug repair, precise binary analysis, adaptive fuzzing test, reverse engineering, and system log analysis. Tools such as LATTE, Zeroleak, and Repiolt demonstrate their robustness in mitigating security threats, while fine-tuning and contrastive learning techniques enhance bug detection precision. Moreover, LLM-driven fuzz testing and binary analysis optimize vulnerability assessment methodologies, in addition, their integration into system log analysis facilitates real-time anomaly detection and proactive cyber threat response. \end{takeaway}
\subsubsection{Vulnerability Detection} The growing number of reports in Common Vulnerabilities and Exposures (CVEs) highlights the rise in software vulnerabilities, increasing the risk of breaches, and posing major economic and social threats. As such, detecting vulnerabilities has become a critical requirement for protecting software systems and ensuring societal and economic stability.

Recent advances have showcased the potential of LLMs in vulnerability detection tasks, particularly for static code analysis, where they demonstrate superior performance compared to traditional methods such as graph neural networks or rule-based approaches~\cite{quan2023,thapa2022,ullah2024}. The GPT series of models stand out in their ability to identify vulnerabilities effectively \cite{khare2024,liu-etal-2023,ullah2024,zhang2024b}. However, challenges persist, as LLMs can generate false positives due to subtle variations in function names, variable usage, or library modifications.
Notable advances include LATTE, proposed by Liu~\etal \cite{liu2024b}, which combines LLMs with automated binary taint analysis. LATTE addresses the limitations of traditional taint analysis methods that rely heavily on manual customization of taint propagation and vulnerability inspection rules. Impressively, LATTE identified 37 previously undiscovered vulnerabilities in real firmware, showcasing the practical impact of LLMs in vulnerability detection.
Tihany~\etal\cite{TihanyiBJFCM23} further demonstrated the utility of LLMs by using them to generate FormAI, a large-scale vulnerability labeled dataset. However, their study also revealed a critical concern: more 50\% of the code generated by LLMs contained vulnerabilities, raising significant security risks for automated code generation processes. These findings underscore the dual-use nature of LLMs, highlighting both their potential to advance vulnerability detection and the need for rigorous oversight to mitigate associated risks.

\subsubsection{Vulnerability Repair} The rapid increase in detected vulnerabilities, coupled with the increasing complexity of modern software systems, has made manual vulnerability remediation an extremely time-consuming and resource-intensive task for security professionals \cite{zhang2024c}. 
Advancements in LLMs and related architectures have demonstrated promising capabilities in automating vulnerability repair tasks. The T5 model, built on an encoder-decoder framework, has shown superior performance in generating effective fixes for vulnerabilities \cite{Fu,zhang2023}. However, challenges persist in maintaining the functional correctness of repaired code \cite{pearce2022}, with LLM performance varying between programming languages, particularly with limited capabilities observed in repairing Java vulnerabilities \cite{Wu_2023}. 
Several innovative approaches have emerged to address these challenges. Alrashedy~\etal\cite{alrashedy2024} developed an automated vulnerability repair tool that integrates feedback from static analysis tools, enabling iterative improvement of fixes. Tol~\etal\cite{tol2023} proposed ZeroLeak, a technique that uses LLM to identify and mitigate side-channel vulnerabilities in applications, demonstrating the capabilities of the models to address complex security challenges.  Charalambous~\etal\cite{CharalambousTJSFC23} combined LLMs with Bounded Model Checking (BMC) to ensure the accuracy of the corrected code, mitigating functionality issues often seen after automated vulnerability fixes.

\subsubsection{Bug Detection}
Software and hardware failures, commonly termed bugs, can cause program malfunctions or unexpected results. Beyond affecting performance, some bugs can be manipulated by attackers to create security vulnerabilities, highlighting the critical need for bug detection to maintain software system safety and reliability. LLMs have emerged as effective tools for automating bug detection. They can generate code lines relative to the original to detect potential bugs. LLMs also utilize feedback from static analysis tools, enhancing the accuracy and precision of bug identification \cite{jin2023,li2023}. Fine-tuning is crucial for adapting LLMs to bug detection, allowing error identification without test cases by using annotated datasets \cite{Lee,yang2023}. Du~\etal\cite{DuY23} and Li~\etal\cite{li2023nu} use contrastive learning to train LLMs to differentiate between correct and faulty code, increasing error detection in complex codebases. Fang~\etal\cite{Fang2023} introduced Represent Them All (RTA), a platform-independent representation method that combines contrastive learning and custom fine-tuning. This technique excels in bug detection and predicts bug priority and severity, highlighting the potential of comprehensive representation methods for software quality.

\subsubsection{Bug Repair} LLMs efficiently automate bug fixes by generating precise code, improving development speed but risking security concerns like vulnerabilities \cite{Perry_2023}. Addressing unresolved bugs is crucial and requires automated repairs in modern software engineering. LLMs are adept at creating repair patches for software defects. Architectures, such as Repilot \cite{Wei_2023}, use encoder-decoder frameworks for accurate repairs, leveraging LLMs' grasp of code semantics to produce high-quality patches on par with traditional techniques \cite{Xia_2024}. Fine-tuning enhances LLMs' real-world repair capabilities with domain-specific datasets for better language and task handling, delivering reliable fixes. Interactive feedback systems, such as ChatGPT \cite{xia2023}, further refine repair precision, supporting effective patch development through iterative validation and a deep understanding of software semantics.

\subsubsection{Program Fuzzing}
 Program fuzzing, fuzz testing, or simply fuzzing, is an automated software testing technique that generates input to uncover unexpected behaviors, such as crashes. Various effective fuzzing tools have successfully detected bugs and security flaws in real systems \cite{Boehme2020}. The incorporation of LLMs into fuzzing has significantly improved the generation of test cases. Traditional methods often rely on predefined patterns, which limits their effectiveness. In contrast, LLMs can produce diverse and contextually suitable test cases for different programming languages and system features \cite{deng2023}. LLMs employ advanced strategies, such as repetitive and iterative querying \cite{Zhang_2024}, to improve the creation of test cases. These methods allow LLMs to create test cases that:
\begin{itemize}
    \item Identify vulnerabilities: LLMs can analyze previous bug reports to create inputs that uncover similar issues in new or updated systems \cite{deng2023l}.
    \item Create variations: They generate various test cases related to sample inputs, ensuring coverage of potential edge cases \cite{HuHIT023}.
    \item Optimize compilers: By examining compiler code, LLMs craft programs that trigger specific optimizations, revealing compilation process flaws \cite{Yang_2024}.
    \item Divide testing tasks: A dual-model interaction lets LLMs separate tasks like test case generation and requirements analysis for efficient parallel processing.
\end{itemize}
The adaptability of LLMs to create intelligent and tailored test inputs has transformed fuzz testing, making it more effective at finding complex bugs. As these models advance, integrating feedback loops, real-time testing, and domain-specific knowledge will further improve system security and robustness.

\subsubsection{Reverse Engineering and Binary Analysis} Reverse engineering involves analyzing artifacts such as software or hardware to discern their functionality, which can be used defensively or maliciously. It is crucial for security in vulnerability analysis, malware investigation, and intellectual property protection.
Although LLMs excel at automating reverse engineering by identifying software functions and extracting essential data, Xu~\etal\cite{xu2024} showcased their ability to restore variable names from binaries through iterative query propagation. Moreover, Armengol~\etal \cite{armengolestapé2024} paired type inference engines with LLMs to disassemble executables and generate source code, simplifying the binary-to-source translation process.
LLMs are crucial in binary program analysis, improving comprehension of low-level code structures and behaviors. Significant progress includes:
\begin{itemize}
    \item DexBert: Proposed by Sun~\etal\cite{sun2023}, this tool characterizes the binary bytecode of the Android system, improving the specific binary analysis of the Android ecosystem.
    \item SYMC Framework: Developed by Pei~\etal\cite{pei2024}, this framework uses group theory to preserve the semantic symmetry of the code during analysis. The approach has demonstrated exceptional generalization and robustness in diverse binary analysis tasks.
    \item Authorship Analysis: Song~\etal\cite{song2022} applied LLMs to address software authorship analysis challenges, enabling effective organization level verification of Advanced Persistent Threat (APT) malicious software.
\end{itemize}
 LLMs also improve the readability and usability of decompiler outputs, helping reverse engineers interpret and understand binary files more effectively. By improving decompiler-generated code, these models reduce manual effort and increase the efficiency of reverse engineering processes.
As LLM technologies advance, their integration with reverse engineering tools is expected to further enhance functionality. Future directions may include real-time disassembly, automated malware deobfuscation, and dynamic binary analysis using hybrid techniques that combine LLMs with symbolic execution or formal verification methods. 

\subsubsection{Malware Detection} The increasing complexity and volume of malware require advanced detection approaches. Signature-based and heuristic methods often fall short against novel camouflaged malware because of the sophisticated evasion techniques used by attackers, such as encryption, polymorphism, and metamorphism. LLMs are effective tools for detecting semantic and structural malware features, thus boosting detection. AVScan2Vec technique proposed by Joyce~\etal\cite{JoycePNR23}, converts antivirus scan reports into vectors, enabling efficient handling of large malware datasets and excelling in tasks like classification and clustering. Using semantic patterns in antivirus data, this method improves scalability and accuracy, presenting a new approach to malware analysis. LLMs have been explored for their role in both detecting and analyzing malware development and prevention. As noted by Botacin \cite{Botacin2023}, LLMs can combine functionalities to create modular malware components, helping to evolve malware variants. Although LLMs cannot independently generate complete malware from prompts, their ability to create elements supports countermeasure development, highlighting the importance of responsible LLM usage to avoid misuse.

\subsubsection{System Log Analysis}Examining extensive log data from software systems manually is impractical due to its complexity and volume. Deep learning techniques have been proposed for anomaly detection within logs. They face challenges such as managing high dimensional, noisy data, tackling class imbalances, and achieving generality \cite{Jakub2017}. Recently, researchers have used the advanced language comprehension capabilities of LLMs to enhance anomaly detection in system logs. LLMs surpass conventional deep learning models in both accuracy and interpretability \cite{Shan_2024}. Their adaptability can be further optimized by fine tuning for specific types of logs \cite{karlsen2023} or by adopting strategies based on reinforcement learning \cite{han2023}, which enables precise identification of anomalies specific to particular domains, LLMs prove beneficial in analyzing cloud server logs \cite{LiuWXQLC23}. By integrating reasoning with log data, they effectively deduce the root causes of issues within cloud services. This demonstrates the impact of LLMs in the realm of system log analysis, providing a scalable and intelligent approach to detect and address anomalies in intricate environments.

\begin{takeaway}
\BfPara{Challenges and Open Directions} Despite their advantages, LLMs face key challenges, including a high false positive rate in detection and functional correctness in automated bug repair. Research indicates that 50\% of LLM-generated code contains exploitable threats, raising concerns. Additionally, scalability challenges in high-throughput environments and ethical risks associated with the dual use of AI necessitate strict oversight and policies. Future research should focus on developing real-time security techniques, improving adversarial defenses, and integrating hybrid AI models with formal verification and reinforcement learning to enhance LLM-driven security solutions' security, reliability, and interpretability. \end{takeaway}

\subsection{ LLMs in Information and Content Security} With phishing, misinformation, manipulation, and cybercrime, LLMs are powerful for enhancing information and content security through context-aware learning, fine-tuned detection models, and advanced response mechanisms that lead to enabling real-time threat identification, fraud prevention, and content moderation with greater accuracy and scalability. This section explores how LLMs redefine information and content security across distinct tasks including phishing and scam detection, harmful content identification, steganography, access control, and digital forensics. It also addresses challenges such as bias, adversarial risks, and ethical concerns.
 
\begin{takeaway}
 LLMs have transformed information and content security through prompt-based learning and fine-tuned models. They accurately detect phishing attempts and scams while proactively disrupting fraud through automated scam engagement. In content moderation and online safety, LLMs enhance the identification of harmful misinformation. In steganography, they facilitate covert data embedding and advance steganalysis using few-shot learning and natural language ciphertext encoding for secure communication. LLMs also support file identification, incident response, and evidence extraction in digital forensics. Tools like PassGPT strengthen authentication by generating high-entropy passwords and evaluating their strength. \end{takeaway}

\subsubsection{Phishing and Scam Detection}
Network deception involves intentionally adding false or misleading information, which threatens user privacy and property security. Typical attack methods include emails, SMS, and web ads that are used to direct users to phishing sites or harmful links \cite{thapa2022}. LLMs can produce deceptive content on a large scale with certain prompts. However, LLMs-generated phishing emails usually have lower click-through rates than manually created ones, highlighting the limitations of automated methods \cite{heiding2023}.
LLMs are highly effective at identifying phishing emails. They use prompt-based strategies with website data or fine-tuned models suited to email traits to reach high effectiveness in phishing detection. They also excel at spotting spam, which often includes phishing. Labonne~\etal \cite{LabonneM23}, show that LLMs significantly outperform traditional machine learning in spam detection, confirming their prowess in this area.
Beyond detection, LLMs offer innovative uses against scams. As shown in \cite{cambiaso2023}, LLMs can mimic human interactions with scammers automatically, wasting their time and resources. This reduces the efficiency of scammers and lessens the impact of scam emails. These abilities show the potential of LLMs in enhancing phishing and scam detection, providing scalable, intelligent, and proactive cybersecurity defenses.

\subsubsection{Harmful Contents Detection}
Social media platforms face criticism for worsening polarization and weakening public discourse. Harmful content, often mirroring users' political opinions, can lead to toxic discussions and harmful behavior. LLMs help detect harmful content in three main areas: identifying extreme political positions \cite{hanley2024}, monitoring crime-related discourse \cite{HuHIT023}, and identifying fake social media accounts or bots \cite{cai2024}.
Although LLMs can identify this content, their interpretations often reflect their internal biases, highlighting the complexities of dealing with intricate social and political topics. Martin~\etal\cite{Martin23} significantly contributed by creating a large-scale dataset of harmful and benign discourse for 13 minority groups using LLM. Validation showed that human annotators struggled to distinguish between LLM-generated and human-authored discourse, indicating LLMs' potential to improve harmful content detection, aiding efforts against toxic behavior.

\subsubsection{Steganography} Anderson \cite{Anderson} described the embedding of secret data within regular information carriers to ensure that hidden content remains secure. This is crucial for secure communication. Recent advancements use LLMs to improve steganography and steganalysis techniques. Wang~\etal\cite{Wang2023} presented a novel steganalysis method using LLMs and few-shot learning. By integrating small labeled datasets with auxiliary unlabeled data, this technique addresses the scarcity of labeled samples, greatly improving detection in low-data scenarios. This marks a significant advancement in language-based steganalysis.
Bauer~\etal\cite{BauerHMBSH21} concurrently showed how the GPT-2 model could encode ciphertext into natural language cover texts. This feature allows users to determine the outward appearance of the ciphertext, making it possible to discreetly transfer sensitive information across public forums. Such developments show the contributions of LLMs in contemporary steganography, offering secure techniques for embedding data, as well as improved methods for detecting potential misuse.

\subsubsection{Access control} Access control is crucial for cybersecurity, designed to limit the actions of authorized users within a system. Despite the new authentication technologies, passwords are still the primary method of enforcing access control \cite{inproceedings}. PassGPT, an advanced password generation system that uses LLMs, presents an innovative strategy for crafting passwords that adhere to user-specified constraints. This technique excels beyond conventional methods, including those based on Generative Adversarial Networs (GANs), by generating a more diverse collection of unique passwords. Furthermore, PassGPT improves the effectiveness of password strength evaluators, underscoring the promise of LLMs to pioneer and strengthen access control measures \cite{rando2023}.

\subsubsection{Forensics} Digital forensics play a crucial role in cybercriminal prosecution by ensuring that evidence extracted from digital devices is allowed in court \cite{Selamat}. This domain protects the integrity and enhances the effectiveness of cybercrime investigations. Scanlon~\etal\cite{ScanlonBHHS23} conducted an evaluation of LLMs within the domain of digital forensics, focusing on tasks such as file identification and responding to incidents. The research concluded that, although LLMs should not be considered independent tools, they offer valuable assistance in particular forensic situations.
\begin{takeaway}
\BfPara{Challenges and Open Directions} Despite their advantages, LLMs face challenges in information and content security. In phishing and scam detection they struggle to replicate human crafted phishing strategies, requiring adversarial training and adaptive threat response. In harmful content detection, LLMs require bias mitigation and a context-aware model to improve accuracy. Their role in steganography enables covert data embedding, increasing the need for advanced steganalysis. In access control, they demand srtonger authentication frameworks. While in digital fornsics,  stronger datasets and legal frameworks are needed. Ensuring ethical, unbiased, and secure deployment remains critical in this domain. \end{takeaway}
\subsection{ LLMs in Hardware Security} SoC architectures, essential for modern computing, integrate multiple IP cores but introduce security challenges, as a weakness in any single core can compromise the entire system. Although software and firmware updates can address many issues, some vulnerabilities cannot be patched this way, necessitating rigorous security measures during the initial design phase. This section provides an overview of LLM applications in hardware security, with a particular focus on their role in detecting and mitigating vulnerabilities. Through these capabilities, LLMs demonstrate their potential to enhance the security of system-on-chip (SoC) architectures.
\begin{takeaway}
 LLMs enhance hardware security by automating vulnerability detection, security verification, and repair mechanisms in SoC architectures. Using NLP driven analysis, they identify threats within hardware design documents and link weaknesses to Common Weakness Enumerations (CWEs) while enforcing security assertions. Additionally, LLMs enhanced hardware vulnerability repair by generating secure hardware code and leveraging large vulnerability corpora to improve patching automation. These advancements make LLMs powerful, scalable, and proactive solutions to strengthen SoC security.  \end{takeaway}
\subsubsection{Hardware Vulnerability Detection} LLMs are being increasingly used to detect hardware vulnerabilities by scrutinizing security aspects embedded in hardware development documents. In an illustration of their capabilities, Meng~\etal\cite{MengSARHP23} leveraged HS-BERT, a model trained on a variety of hardware architecture documents, including Reduced Instruction Set Computing (RISC-V), 
 Open Source Reduced Instruction Set Computing (OpenRISC), and Microprocessor without Interlocked Pipeline Stages (MIPS), facilitated the discovery of eight security vulnerabilities in the OpenTitan SoC design. This demonstrates LLMs' proficiency in dissecting complex hardware configurations to pinpoint significant security issues. Building on this advancement, Paria~\etal\cite{paria2023} extended the utility of LLMs by identifying vulnerabilities within user-specified SoC designs. Their innovative strategy links identified vulnerabilities to CWEs, makes corresponding security assertions, and enforces security measures to counteract potential threats. These cutting-edge advances demonstrate the importance of LLMs in improving hardware security by streamlining the process of detecting and averting vulnerabilities in advanced hardware infrastructures.

\subsubsection{Hardware Vulnerability Repair} LLMs play an essential role in the security verification of SoC. They handle a variety of tasks, including the insertion of vulnerabilities, their assessment, and verification, as well as the development of strategies for their mitigation~\cite{saha2023}. By utilizing comprehensive data on hardware vulnerabilities, LLMs provide suggestions for repairs, which significantly enhance the effectiveness and precision of security assessments and mitigation efforts.
Nair~\etal\cite{Nair2023} demonstrated that LLMs can identify hardware vulnerabilities while generating code and can produce hardware code that prioritizes security. In their study, they utilized LLMs to design hardware that effectively addresses ten identified CWEs. In a complementary study, Tan~\etal\cite{Ahmad_2024} constructed an extensive corpus detailing hardware security vulnerabilities and evaluated the proficiency of LLM in automating the repair of these vulnerabilities. 

\begin{takeaway}
\BfPara{Challenges and Open Directions} LLMs enhance hardware vulnerability detection and repair, but face challenges in accurately interpreting complex SoC architectures, which require deep contextual understanding. The generalization of LLMs across diverse hardware designs is a limitation, as models trained on specific architectures may not effectively identify vulnerabilities in unfamiliar systems. Additionally, linking detected vulnerabilities to CWEs and generating security assertions require further refinement to ensure precision and reliability in automated mitigation efforts. Open research directions include developing specialized LLMs tailored for hardware security, improving adaptive security assertion generation, and integrating LLMs with formal verification methods to improve the accuracy of security enforcement. \end{takeaway}

\subsection{LLMs in Blockchain Security} 

Blockchain technology has revolutionized decentralized finance, digital identity, and secure transactions by providing a transparent ledger system. However, its growing adoption has also introduced critical security challenges, particularly in smart contract vulnerabilities and transaction anomalies. As blockchain ecosystems become increasingly complex, the need for intelligent, scalable, and proactive security mechanisms has become essential. LLMs present a transformative solution. In this section, we explore how LLMs impact blockchain security, particularly in smart contract security and the identification of transaction anomalies. Examining their substantial capabilities highlights the potential of LLMs to transform vulnerability management, thereby strengthening blockchain systems by mitigating risks and enhancing overall security.

\begin{takeaway}
 LLMs enhance blockchain security by enabling smart contract vulnerability detection and real-time transaction anomaly identification. Frameworks like GPTLENS employ a dual-phase approach to generate and prioritize threat scenarios, reducing false positive rates and improving verification accuracy. Unlike rule-based models, LLMs dynamically identify abnormal transactions without predefined constraints, allowing for a broader spectrum of anomalies, improving intrusion detection efficiency, and minimizing the need for manual analysis.  \end{takeaway}

\subsubsection{Smart Contract Security} Blockchain applications are heavily based on smart contracts, but the construction of these contracts can result in vulnerabilities that pose significant risks, including the potential for substantial financial loss. LMs offer the potential to automate the detection of these vulnerabilities, though their performance is frequently hampered by common errors and a limited understanding of context \cite{Chen,david2023}. To address these challenges, frameworks such as GPTLENS \cite{HuHIT023} employ a dual-phase approach. This process begins with the creation of a wide range of potential vulnerability scenarios and then proceeds with the evaluation and prioritization of these scenarios to minimize false positive detections. Sun~\etal\cite{SunWXLW24} contributed to the field of smart contract security by incorporating LLMs into the analysis of the program to detect logical vulnerabilities effectively. They structured their approach by categorizing vulnerabilities into specific scenarios and attributes, using LLMs for detection, and confirming findings with static analysis. This development shows the ability of LLMs to improve smart contract security, but there remains a need to reduce false positives and increase accuracy.

\subsubsection{Transaction Anomaly Detection} Real-time detection of intrusions within blockchain transactions is challenging due to the extensive search space and the substantial amount of manual analysis required. Conventional methods, such as reward-based and pattern-based models, rely on designated rules or predetermined patterns to pinpoint profitable or suspicious transactions. However, these methods often fail to detect a wide range of anomalies \cite{RodlerRLO19}.
LLMs offer a versatile and generalizable solution for real-time detection of anomalies. Gai~\etal\cite{LuoLV24} demonstrated that LLMs are capable of dynamically identifying anomalies in blockchain transactions as they occur. Unlike traditional methods, LLMs are not constrained by predefined rules or limited search spaces, which enables them to detect a broader array of abnormal transactions. This flexibility highlights the potential of LLMs to advance anomaly detection in blockchain, delivering more efficient and comprehensive solutions.

\begin{takeaway}
\BfPara{Challenges and Open Directions} Despite advancements in LLMs for blockchain security, they still face challenges such as contextual limitations, false positives, and limited understanding of contract logic. Scalability, computational efficiency, and generalization among blockchain protocols also remain significant hurdles. Future research should focus on improving the contextual reasoning of LLMs, integrating formal verification, and developing hybrid AI-driven security frameworks that combine symbolic execution, reinforcement learning, and deep learning-based anomaly detection to improve the accuracy and robustness of LLMs in blockchain security. \end{takeaway}

\subsection{LLMs in Cloud Security} 

The dynamic nature of cloud environments requires real-time threat intelligence, automated security enforcement, and proactive anomaly detection to ensure system integrity and data protection. The integration of LLMs into cloud security has significantly improved threat detection, security monitoring, and data leakage prevention. By employing advanced NLP techniques, these models have improved automation and overall efficiency in addressing complex security challenges, such as misconfigurations, data leaks, compliance issues, and container security.

\begin{takeaway}
 LLMs have transformed cloud security by enhancing threat detection, misconfiguration analysis, data leakage monitoring, container security, and compliance enforcement. In misconfiguration detection, tools like GenKubeSec provide automated reasoning and high-precision detection in Kubernetes environments. Data leakage monitoring benefits from AI-driven models such as Secure Cloud AI, which improves real-time detection. Frameworks like DevSec-GPT enhance vulnerability tracking and compliance validation. Additionally, OllaBench and PRADA demonstrate the effectiveness of LLMs by automating regulatory compliance and integrating Zero Trust security models across multi-cloud environments. \end{takeaway}

\subsubsection{Misconfiguration Detection}Misconfiguration detection plays an essential role in ensuring the security and stability of systems in cloud native settings. Recent innovations have incorporated machine learning and LLMs to effectively identify, pinpoint, and address these misconfigurations. A noteworthy advancement was made by Mitchel~\etal\cite{mitchell2024}, who developed a tool designed to utilize system call data obtained from Linux kernels operating within Kubernetes clusters. This tool applies anomaly detection techniques, including Principal Component Analysis (PCA), to detect hidden attacks that capitalize on misconfigurations. Pranata~\etal\cite{pranata2020} proposed a framework integrating metamorphic testing with PCA to identify misconfigurations in cloud-native applications, enhancing scalability and reducing developer effort. Malul~\etal\cite{MalulMMES24} developed GenKubeSec, an LLM-based system effective in detecting Kubernetes configuration file misconfigurations, offering automated reasoning and solutions. GenKubeSec achieved a precision of 0.990 and a recall of 0.999, surpassing traditional rule-based tools and using a UMI for standardized evaluations.

\subsubsection{Data Leakage Monitoring} Data leakage poses a significant threat to security in cloud computing by compromising the confidentiality and integrity of sensitive information. Issues typically stem from misconfigured hypervisors, inadequate dashboard authentication, and insecure VM replication, especially during operations like data migration.
Ariffin~\etal\cite{ariffin2019} recommended using Wireshark for packet analysis to monitor data flows during VM migration and dashboard authentication, revealing significant risks in the lack of TLS encryption, which exposed credentials. Vaidya~\etal\cite{Vaidya2016} suggested perturbation and fake object injection techniques to better detect unauthorized access by embedding decoy elements. Advanced AI-driven frameworks are revolutionizing data leakage monitoring. Kasula~\etal\cite{kasula2024} introduced ``Secure Cloud AI,'' a hybrid model combining Random Forest and LSTM networks for real-time anomaly detection. Their method achieved 94.78\% accuracy in identifying malware and efficiently classifying network traffic anomalies, demonstrating the necessary scalability and adaptability for dynamic cloud settings. However, challenges persist in scaling for large cloud installations, detecting complex multi-layered attacks, and achieving cross-platform compatibility. Future work should enhance AI integration, real-time encryption, and anomaly detection to strengthen defenses against data leakage.

\subsubsection{Container Security}Integrating LLMs into container security has shown significant progress in securing native cloud environments by automating vulnerability detection, optimizing container management, and improving pipeline integrity. For example, the DevSec-GPT~\cite{Bandara2024} framework uses Meta's Llama2 LLM to create Pipeline Bills of Materials (PBOMs) for container security, analyzing vulnerabilities and development data to ensure comprehensive tracking and prevent supply chain attacks via blockchain traceability. In runtime and security management, LLMs improve real-time anomaly detection by analyzing logs and configurations. Lanka~\etal\cite{lanka2024} employed LLMs to analyze data from the decoy system, detecting malicious patterns and attacker tactics. The RAG model quickly identified threats in containers by matching commands with adversary data. Additionally, LLMs facilitated compliance checks through JSON schema generation for vulnerability scans on platforms such as GitHub Actions and Kubernetes. Automated documentation updates helped meet regulatory standards, while blockchain features, such as NFT tokenization, improved data provenance and auditability.

\subsubsection{Compliance Enforcement} LLMs transform cloud security compliance by automating regulations and boosting efficiency. Nguyen~\etal\cite{ollabench2024} introduced OllaBench, an evaluation tool with 24 cognitive behavioral theories to test the reasoning of LLM in compliance. OllaBench revealed that GPT-4o and Claude are effective in regulatory automation, providing key information to compliance teams.
Henze~\etal\cite{henze2020} proposed PRADA, a cloud storage system using LLMs for transparent data management. It achieves compliance by tagging and routing data by attributes, addressing issues like localization and encryption in distributed systems. PRADA notably improved the management of Data Handling Requirements (DHRs), providing a scalable compliance solution for multi-cloud environments. Dye~\etal\cite{dye2024} highlighted the critical role of LLMs in integrating Zero Trust Architectures (ZTA) with Attribute-Based Access Control (ABAC) systems for compliance. Cloud providers such as AWS utilized LLMs to automate policy generation, monitor privilege escalation, and adhere to Federal Risk and Authorization Management Program (FedRAMP) and National Institute of Standards and Technology (NIST) standards, enhancing least privilege access control and increasing security in hybrid and public clouds.

\begin{takeaway}
\BfPara{Challenges and Open Directions} Despite progress in applying LLMs to cloud security, major challenges persist: scalability in large environments, detecting multilayered attacks, and maintaining cross-platform compatibility. Data leakage prevention demands tighter AI integration, real-time encryption, and adaptive anomaly detection. Container security calls for efficient, real-time anomaly detection and compliance checks. Compliance enforcement remains hindered by real-time auditing, localization, and evolving regulations. Future research should target self-learning models, Zero Trust architectures, and automated security frameworks to improve scalability, adaptability, and resilience.\end{takeaway}

\subsection{LLMs in Incident Response and Threat Intelligence}As cyber threats become more advanced, traditional methods of incident response and threat intelligence often struggle with scalability, speed, and accuracy. LLMs are emerging as transformative tools in incident response and threat intelligence by automating cybersecurity data analysis, enhancing decision making, and improving the speed and accuracy of threat detection. Their advanced pattern recognition allows security teams to identify anomalies and threats within vast amounts of unstructured data, reducing manual effort and response time.

\begin{takeaway}
 LLMs are reshaping incident response and threat intelligence through automation and improved precision. SEVENLLM cuts down false positives in SIEMs via multitask learning for smarter alerting. HunGPT boosts threat analysis with interpretable anomaly detection and structured knowledge extraction. DISASLLM advances malicious code detection, while MALSIGHT generates readable summaries of malware behavior. Together, these tools streamline reverse engineering and speed up intelligence workflows, showing LLMs' growing role in cybersecurity.  \end{takeaway}

\subsubsection{Alert Prioritization}LLMs are revolutionizing alert prioritization in incident response and threat intelligence, supporting efficient context-driven security operations. Molleti~\etal\cite{Molleti2024} showcased the ability of LLM agents to manage large security data sets, alleviate alert fatigue, and highlight key threats through advanced NLP. These agents seamlessly integrate with SIEM systems, boosting threat detection and response. Ji~\etal\cite{ji2024} introduced SEVENLLM, an optimized LLM framework using selected bilingual data. Specializing in analyzing and prioritizing security alerts, SEVENLLM employs multitask learning and SEVENLLM Bench evaluations. This approach improves Indicator of Compromise (IoC) detection and refines alert prioritization by evaluating severity and impact. LLMs significantly improve alert management by reducing false positives and enabling quick responses; however, challenges persist, such as model interpretability and adaptation to evolving threats, underscoring their impact on modern cybersecurity.

\subsubsection{Automated Threat Intelligence Analysis}Incorporating LLMs into cybersecurity has greatly improved automated threat analysis by reducing the manual work involved with unstructured CTI reports. Tseng~\etal\cite{tseng2024} presented an AI agent using LLMs such as GPT-4 to automatically extract IoC from CTI reports and create regex patterns for Security Information and Event Management (SIEM) systems. The agent also constructs relationship graphs to depict connections between IoCs, streamlining incident response, and reducing dependency on human intervention. Using LLMs like Llama 2 and Mistral 7B, Fieblinger~\etal\cite{fieblinger2024} generated Knowledge Graphs (KGs) from CTI reports. Their approach involves fine-tuning and prompt engineering to derive triples, which are employed in link prediction. KGs offer better-structured data representation, enhancing decision making and threat prediction.
HuntGPT, presented by Ali and Kostakos \cite{ali2023}, showcases the capabilities of LLM in cybersecurity. This dashboard combines GPT-3.5 with XAI frameworks such as SHAP and LIME, delivering understandable threat intelligence. It emphasizes detected anomalies and clarifies their context, boosting trust and enabling dynamic cybersecurity processes. These innovations highlight the role of LLM in automating threat tasks, speeding up response, and enhancing detection accuracy in security operations.

\subsubsection{Threat Hunting} LLMs are revolutionizing threat hunting by automating the analysis of intricate cybersecurity data. Schwartz~\etal\cite{schwartz2024} developed LLMCloudHunter, a framework using LLMs like GPT-4o to create OSCTI detection rules. It achieved 92\% precision and 98\% recall, improving rule generation for SIEM systems and boosting threat detection in cloud environments. Mitra~\etal\cite{mitra2024} presented LOCALINTEL, a system that merges global threat intelligence (e.g., CVE, CWE) with local knowledge using RAG, reducing Security Operations Center (SOC) analysts, LLMs achieved a RAGAS score of 0.9535. Their ability to automate threat detection and response promises scalable, accurate, proactive cybersecurity. Future work should improve real-time adaptability and integration with evolving threats.

\subsubsection{Malware Reverse Engineering} The integration of LLMs into malware reverse engineering has transformed the ability to analyze complex binary malware. DISASLLM, introduced by Rong~\etal\cite{rong2024}, leverages an LLM-based classifier fine-tuned in assembly code to efficiently identify valid instruction boundaries within hidden executables. This approach significantly outperforms traditional disassembly tools by integrating the semantic understanding of LLMs, thereby improving the detection of malicious code segments. Complementing this, MALSIGHT, as proposed by Lu~\etal\cite{lu2024}, employs LLMs like MalT5 to iteratively generate human-readable summaries of malware functionality from malicious source code and benign pseudocode. This method improves the usability, accuracy, and completeness of malware behavior descriptions, effectively bridging the semantic gap introduced by obfuscation techniques. Furthermore, Patsakis~\etal\cite{patsakis2024} demonstrated the capabilities of LLMs such as GPT-4 to clarify real-world malware campaigns such as Emotet. The study shows that these models can obtain actionable information, like C2 server configurations, from heavily obfuscated scripts. Although local LLMs have accuracy issues, cloud-based tools like GPT-4 excel in understanding malware payloads. 

\begin{takeaway}
\BfPara{Challenges and Open Directions} 
Despite their influence in this domain, LLMs face key challenges. In alert prioritization, contextual accuracy is required to minimize false positives and misclassifications in SIEM systems. Threat intelligence automation struggles with processing hidden data and adapting to new patterns in real time. Effective threat hunting requires deeper integration with security monitoring data and retrieval-augmented intelligence. Future research must focus on improving contextual LLM reasoning, integrating Explainable Artificial Intelligence (XAI) for transparent threat analysis, and improving real-time adaptability.\end{takeaway}

\subsection{LLMs in IoT Security}
The rapid expansion of IoT ecosystems has introduced significant security challenges due to resource constraints, diverse architectures, and evolving cyber threats. Traditional security solutions struggle with scalability, real-time anomaly detection, and firmware vulnerability management, making IoT devices prime targets for cyberattacks. LLMs have emerged as transformative solutions that enable automated threat detection and efficient processing. The following subsections explore the role of LLMs in enhancing IoT security, focusing on firmware vulnerability detection, behavioral anomaly detection, and automated threat report summarization.

\begin{takeaway}
 
LLMs have significantly advanced IoT security by enhancing firmware vulnerability detection, behavioral anomaly detection, and automated threat report summarization. Frameworks such as LLM4Vuln improve firmware vulnerability analysis by integrating retrieval-augmented generation (RAG) and prompt engineering, enhancing reasoning across various programming languages. The UVSCAN framework employs NLP-driven binary analysis, excelling in detecting API misuse. In behavioral anomaly detection, an Intrusion Detection System Agent (IDS-Agent) combines reasoning pipelines, memory retrieval, and external knowledge to detect zero-day attacks. These advancements integrate LLM reasoning, NLP frameworks, and efficient learning models, providing robust, scalable, and resource-efficient solutions for IoT firmware vulnerabilities.
\end{takeaway}

\subsubsection{Firmware Vulnerability Detection} 
LLMs have notably improved IoT firmware vulnerability detection by translating abstract security needs into actionable analyzes. Sun~\etal\cite{SunWXLW24} introduces LLM4Vuln, which separates the reasoning abilities of LLMs for accurate vulnerability detection in languages like Solidity and Java. The framework improves performance using advanced prompt engineering and RAG to integrate current vulnerability knowledge. Zhao~\etal\cite{Zhao2023} noted that the UVSCAN framework supplements this by translating high-level API specifications into binary-level analysis in IoT firmware using NLP-driven methods. It excels at identifying API misuse, causality errors, and return value issues, offering scalability across various architectures, such as RISC and MIPS. Furthermore, Li~\etal\cite{li2024} introduce Binary Neural Networks (BNNs) to enhance resource efficiency in IoT settings, allowing lightweight on-device learning for vulnerability detection with maintained accuracy. 

\subsubsection{Behavioral Anomaly Detection}With the rapid increase in IoT devices, security challenges have intensified, making the detection of behavioral anomalies vital. LLMs have transformed this field by leveraging reasoning and contextual understanding. Li~\etal\cite{li2024a} introduced IDS-Agent, an intrusion detection system powered by LLMs, which combines reasoning pipelines, external knowledge, and memory retrieval to detect malicious traffic accurately. In benchmarks such as the Army Cyber Institute for Cybersecurity of Things 2023 (ACI-IoT'23) and the Canadian Institute for Cybersecurity Internet of Things 2023 (CIC-IoT'23), IDS-Agent achieved a 61\% recall in identifying zero-day attacks, outperforming traditional machine learning methods and providing better interpretability. Su~\etal\cite{su2024} applied GPT-4o and domain-adapted BERT to IoT time series data, detecting behavioral shifts for threat identification. These models scaled efficiently in resource-limited IoT networks, minimized false alarms, and provided actionable insights to address incidents.

\subsubsection{Automated Threat Report Summarization}The rise of IoT devices has increased the volume and complexity of threat data, emphasizing the need for automated processing. LLMs effectively transform unstructured IoT threat reports into actionable insights. Feng~\etal\cite{feng2019} developed IoTShield, an LLM-based framework for assessing IoT vulnerability reports, extracting critical details such as exploit parameters and severity ratings, and generating custom signatures for IDS to enhance defense accuracy. Building on this, Baral~\etal\cite{baral2024} integrated XAI with LLMs to produce personalized threat reports tailored to the expertise of analysts. Their system balances technical complexity with user-friendliness, improving decision making and response efficiency. These advancements underscore the role of LLMs in optimizing IoT threat intelligence processes and addressing challenges related to scale, complexity, and interpretability in modern IoT security environments.

\begin{table}[h!]
 \centering
\small
\scalebox{0.8}{\begin{tabular}{p{4cm}|p{4cm}|c}
\hline
\textbf{Security Domains}          & \textbf{Security Tasks}                                                                                 & \textbf{Total} \\ \hline
\multirow{1}{*}{Network Security}  & Web fuzzing (3) \newline Traffic and intrusion detection (3) \newline Cyber threat analysis (3) \newline Penetration test (4) & 13            \\ \hline
\multirow{2}{*}{Software and System Security} 
                                    & Vulnerability detection (9) \newline Vulnerability repair (8) \newline Bug detection (7) \newline Bug repair (4) \newline Program fuzzing (6) \newline Reverse engineering and binary analysis (5) \newline Malware detection (2) \newline System log analysis (5) & 46 \\ \hline
\multirow{1}{*}{Information and Content Security}  
                                    & Phishing and scam detection (4) \newline Harmful contents detection (4) \newline Steganography (3) \newline Access control (2) \newline Forensics (2) & 15            \\ \hline
\multirow{1}{*}{Hardware Security} & Hardware vulnerability detection (2) \newline Hardware vulnerability repair (3)                          & 5             \\ \hline
\multirow{1}{*}{Blockchain Security} & Smart contract security (4) \newline Transaction anomaly detection (3)                                   & 7             \\ \hline
\multirow{1}{*}{Cloud Security}  & Misconfiguration detection (3) \newline Data leakage monitoring (3) \newline Container security (2) \newline Compliance enforcement (3) & 11            \\ \hline
\multirow{1}{*}{Incident Response and Threat Intel.}  & Alert prioritization (2) \newline Automated threat intelligence analysis (3) \newline Threat hunting (2) \newline Malware reverse engineering (3) & 10            \\ \hline
\multirow{1}{*}{IoT Security}  & Firmware Vulnerability Detection (3) \newline Behavioral Anomaly Detection (2) \newline Automated Threat Report Summarization (2) & 7     \\ \hline
\end{tabular}}
\caption{Security Domains and Related Tasks}
\label{tab:security_tasks}
 \end{table} 

\autoref{tab:security_tasks} categorizes key cybersecurity domains along with their associated tasks and corresponding counts, highlighting the potential of LLMs to enhance security operations. We examine the applications of LLMs across 32 security tasks spanning eight distinct security domains. Furthermore, this classification serves as a foundational reference for the role of LLMs in cybersecurity research, facilitating more adaptive and intelligent security.

\begin{takeaway}
\BfPara{Challenges and Open Directions} 
Firmware vulnerability detection still requires improved cross-architecture generalization to enhance accuracy in various IoT environments. Behavioral anomaly detection struggles with reducing false positives in complex IoT ecosystems while maintaining efficient resource utilization in constrained environments. Automated summarization of threats and reports demands better contextual understanding to generate actionable insights. Future research should focus on developing energy-efficient LLM models for IoT, creating lightweight architectures, and improving real-time intrusion detection through adaptive learning.\end{takeaway}

\section{Vulnerabilities and Defenses}
\label{sec:Vulnerabilities_and_Mitigation_Strategies}
Existing researches works have classified the vulnerabilities and challenges associated with LLMs into distinct domains. Security and privacy risks include misinformation \cite{pathak2023}, trustworthiness concerns \cite{liu2024}, hallucinations \cite{kaddour2023}, and significant resource consumption \cite{tornede2024}. These risks focus on the need for robust measures to mitigate potential security attacks. Security is mainly intended to protect systems by preventing unauthorized access, modification, malfunction, or denial of service to legitimate users during regular operations \cite{csrc2023}. On the other hand, privacy aims to protect personal information and ensure that individuals retain control over who can access their sensitive data \cite{cloudflare2023}.
In this work, our objective is to systematically examine LLM vulnerabilities by focusing on security attacks through a goal-oriented approach.
The subcategories include Backdoor Attacks, Data Poisoning, Prompt Injection, and Jailbreaking, each paired with the corresponding defense techniques. This structure highlights the challenges and innovative countermeasures in securing LLMs, providing a clear framework for understanding and addressing these risks. The following subsections first explore key primary defense techniques against security attacks on LLMs, followed by an analysis of major security attack types with their defense methods, focusing on a secure and reliable LLM deployment.

 \subsection{Defenses Against Attacks on LLMs}
 This subsection outlines key defense techniques proposed to enhance the robustness and safety of LLMs, belonging to three main categories. The first focuses on preventing LLMs from generating harmful output using a set of rules and constraints at the input or output levels. Red teaming and content filtering play a critical role in intercepting and blocking potentially harmful interactions before they occur, ensuring that LLMs adhere to ethical and safety standards. The second category is concerned with modifying LLMs internal mechanisms or representations to improve LLM robustness and safety. Safety fine-tuning and model merging are key techniques within this category,  working to make the model more resilient to adversarial attacks and misalignments, while strengthening its safety protocols through model optimization. The third category integrates the strengths of these defense strategies to offer a more comprehensive protection framework.
 
 \begin{enumerate}[leftmargin=*]

    \item \textbf{Red Team Defenses.} This is an effective technique for simulating real-world attack scenarios to identify LLM vulnerabilities. The process begins with attack scenario simulation, where researchers test LLM responses to issues such as abusive language. This is followed by test case generation using classifiers to create scenarios that help eliminate harmful outputs. Finally, the attack detection process assesses the susceptibility of LLMs to adversarial attacks. Continuous updates to security policies, refinement of procedures, and strengthening of technical defenses ensure that LLMs remain robust and secure against evolving threats. Ganguli~\etal~\cite{GanguliLKABK22} proposed an efficient AI-assisted interface to facilitate large-scale Red Team data collection for further analysis. Additionally, their research explored the scalability of different LLM types and sizes under Red Team attacks and their ability to reject various threats.

    \BfPara{Challenges} This technique poses several challenges, such as its resource-intensive nature and the need for skilled experts to effectively simulate complex attack strategies~\cite{R_ttger_2021}. Red Teaming is still in its early stages with limited statistical data. However, recent advancements, such as leveraging automated approaches for test case generation and classification, have improved scalability and diversity in Red Teaming efforts~\cite{perez2022, ganguli2022}.

    \item \textbf{Content Filtering.} This technique encompasses input and output filtering to protect the integrity and appropriateness of LLM interactions by identifying harmful inputs and outputs. Recent advancements have introduced two key approaches: rule-based and learning-based systems. Rule-based systems rely on predefined rules or patterns, such as detecting adversarial prompts with high perplexity values~\cite{alon2023}. To neutralize semantically sensitive attacks, Jain~\etal~\cite{jain2023} utilize paraphrasing and re-tokenization techniques. On the other hand, learning-based systems employ innovative methods, such as an alignment-checking function designed by Cao~\etal~\cite{CaoXHZ2022} to detect and block alignment-breaking attacks, enhancing security.

    \BfPara{Challenges} Despite advancements, this technique still suffers from significant limitations that hinder its effectiveness and robustness. One key issue is the evolving nature of adversarial prompts designed to bypass detection mechanisms. Rule-based filters, while simple, struggle to remain effective against increasingly sophisticated attacks. Learning-based systems, which rely on large and diverse datasets, may fail to fully capture harmful content variations and struggle with balancing sensitivity and specificity. This imbalance can lead to false positives (flagging benign content) or false negatives (missing harmful content). Additionally, the lack of explainability in machine learning-based filtering decisions complicates transparency and acceptance, underscoring the need for more interpretable and trustworthy models.

    \item \textbf{Safety Fine-Tuning.} This widely used technique customizes pre-trained LLMs for specific downstream tasks, offering flexibility and adaptability. Recent research, such as that presented by Xiangyu~\etal~\cite{qi2024}, found that fine-tuning with a few adversarially designed training examples can compromise the safety alignment of LLMs. Additionally, even benign, commonly used datasets can inadvertently degrade this alignment, highlighting the risks of unregulated fine-tuning. Researchers have proposed data augmentation and constrained optimization objectives to address these challenges. Data augmentation enriches the training dataset with a diverse range of samples, including adversarial and edge cases, helping the model generalize safety principles more effectively. Constrained optimization applies additional loss functions or restrictions during training to guide the model toward prioritizing safety without sacrificing task performance. Bianchi~\etal~\cite{BianchiSA24} and Zhao~\etal~\cite{ZhaoDMZR24} validate the effectiveness of this technique, suggesting that incorporating a small number of safety-related examples during fine-tuning improves the safety of LLMs without reducing their practical effectiveness. Together, these defensive techniques enhance safety throughout the generative process, reducing the risk of adversarial attacks and unintentional misalignments while preserving the adaptability and effectiveness of fine-tuned LLMs for real-world tasks.

    \BfPara{Challenges} While widely used, integrating safety-related examples into fine-tuning can limit the generalization of LLMs to benign, non-malicious inputs or degrade performance on specific tasks. Additionally, the reliance on manually curated safety examples or prompt templates demands considerable human expertise and is subject to subjective biases, leading to inconsistencies and affecting model reproducibility across different use cases. Moreover, fine-tuning itself can both enhance safety through the introduction of safety-focused examples and expose the model to new vulnerabilities when adversarial examples are incorporated.

    \item \textbf{Model Merging.} The model merging technique combines multiple models to enhance robustness and improve performance~\cite{wortsman2022}. This method complements other defense strategies by significantly strengthening the resilience of LLMs against adversarial manipulations. By leveraging the diversity of multiple models, it creates a more robust system capable of handling adversarial inputs while maintaining generalization across various tasks. When merging models, different fine-tuned models initialized from the same pre-trained backbone share optimization trajectories while diverging in specific parameters tailored to different tasks. These diverging parameters can be merged through arithmetic averaging, allowing the model to generalize better over domain inputs and perform multi-task learning. This idea has been proven effective in fields like Federated Learning (FL) and Continual Learning (CL), where model parameters from different tasks are combined to mitigate conflicts. Zou~\etal~\cite{zou2023} and Kadhe~\etal~\cite{kadhe2024} applied model merging techniques to balance unlearning unsafe responses while minimizing over-defensiveness.

    \BfPara{Challenges} Despite its potential, model merging faces significant challenges due to a lack of deep theoretical investigation in two key areas. First, the relationship between adversarially updated model parameters derived from unlearning objectives and the embeddings associated with safe responses remains unclear. There is a risk that adversarial training with a limited set of harmful response texts could lead to overfitting, making the model more susceptible to new, unseen jailbreaking prompts. This limited approach may fail to generalize effectively to novel adversarial threats, undermining the robustness of LLMs. Second, controlling the over-defensiveness of merged model parameters presents a significant challenge. While merging aims to improve resilience, it does not provide a clear methodology for preventing over-defensive behavior, where the model may excessively restrict certain types of outputs, limiting its usability and flexibility. This lack of control could hinder the model's ability to balance safety with task performance, as overly defensive behavior may result in missed or overly cautious responses, impacting user experience and task accuracy. These challenges highlight the need for further research into the theoretical foundations of model merging to ensure its effectiveness and versatility as a defense mechanism.

\end{enumerate}

 \subsection{Adversarial Attacks} Adversarial attacks are a key vulnerability in LLMs, involving input manipulation to trigger errors or unintended output \cite{zhang2024c}. These attacks exploit the sensitivity of a model to minor input changes to deceive it. In DNNs, such attacks disrupt operations by altering input data, and this also applies to LLMs, with potential effects such as spreading misinformation or creating biased content \cite{xu2023}.

\begin{figure}[t]
    \centering
    \includegraphics[width=.8\linewidth]{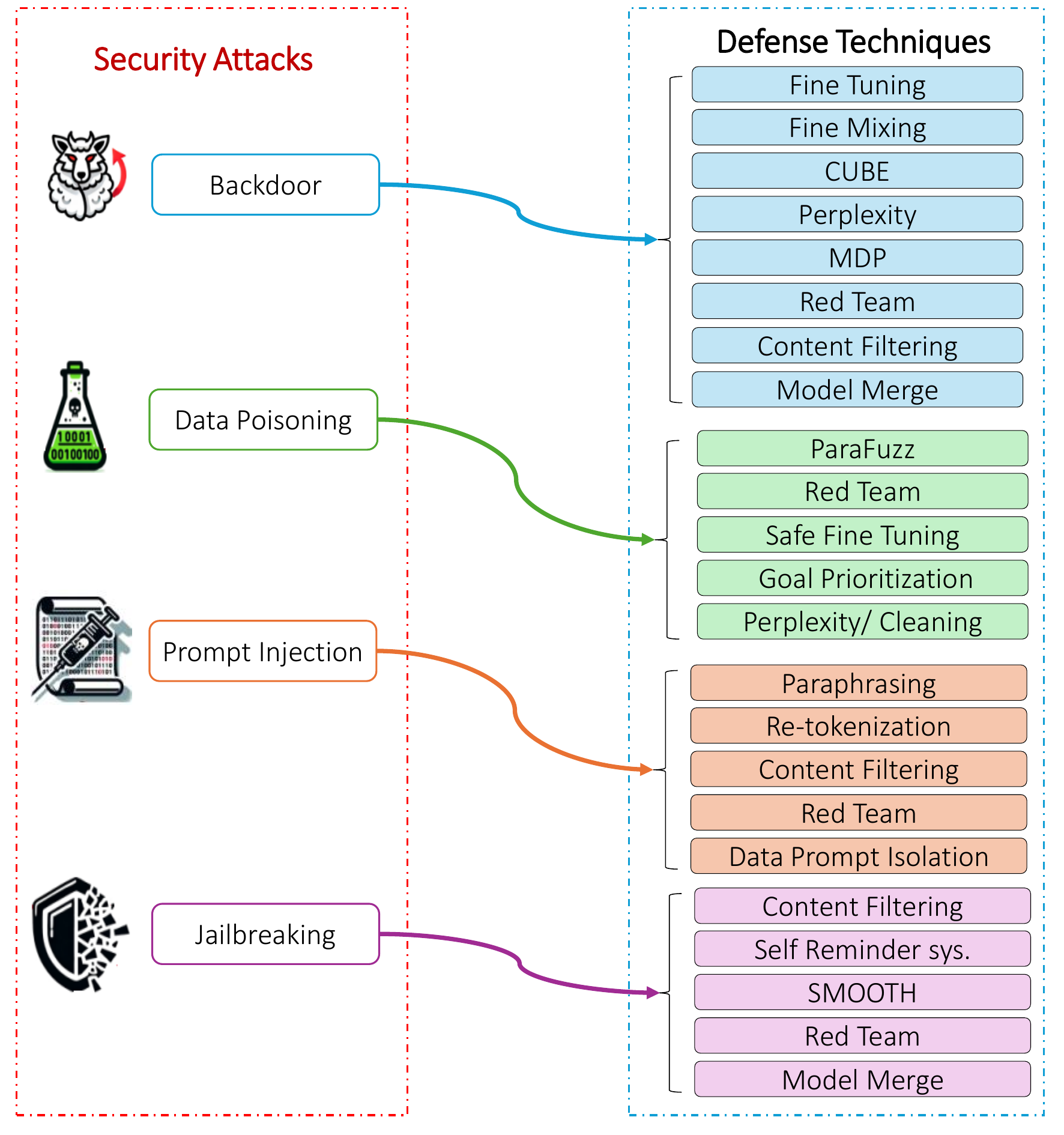}
    \caption{Security Attacks and Defense Techniques}
    \label{Figure three}
\end{figure}

 In the following, we review those attacks and defenses.  At a high level, \autoref{tab:llm_defense} highlights various defense techniques proposed to mitigate LLMs against four key security vulnerabilities (backdoor attacks, jailbreaking, data poisoning, and prompt injection). The authors below explore these techniques, which are used to safeguard LLMs from security attacks by addressing their limitations and challenges. In more detail, \autoref{Figure three} presents a structured overview of security attacks targeting LLMs alongside corresponding defense techniques. By integrating these advanced security frameworks, LLMs can achieve enhanced robustness and reliability for ensuring safe and trustworthy deployment in real-world applications.

     \begin{table*}[t]
    \centering
    \setlength{\tabcolsep}{4pt}
   \scalebox{0.9}{ \begin{tabular}{lcccc} 
        \toprule
        \textbf{Technique} & \textbf{Backdoor} & \textbf{Jailbreaking} & \textbf{Data Poisoning} & \textbf{Prompt Injection} \\
        \midrule
        ParaFuzz       & \ding{53} & \ding{53} & \cmark & \ding{53} \\
        CUBE      & \cmark & \ding{53} & \ding{53} & \ding{53} \\
        Masking Differential Prompting   & \cmark & \ding{53} & \ding{53} & \ding{53} \\
        Self Reminder System   & \ding{53} & \cmark & \ding{53} & \ding{53} \\
        Content Filtering       & \cmark & \cmark & \cmark & \cmark \\
        Red Team           & \cmark & \cmark & \cmark & \cmark \\
         Safety Fine-Tuning             & \cmark & \ding{53} & \cmark & \ding{53} \\
        A Goal Prioritization     & \ding{53} & \cmark & \ding{53} & \ding{53} \\
         Model Merge        & \cmark & \ding{53} & \cmark & \ding{53} \\
        Prompt Engineering     & \cmark & \cmark &\cmark &\cmark \\
         Smooth     & \ding{53} & \cmark &\ding{53} &\ding{53} \\
        \bottomrule
    \end{tabular}}
    \caption{Comparison of Techniques for Mitigating LLM Vulnerabilities}
    \label{tab:llm_defense}
\end{table*}

\subsubsection{Data Poisoning} 
Adversaries pose a serious threat to LLMs by deliberately altering training datasets. Through the insertion of misleading samples, they create subtle distortions that bias the model, leading to errors in prediction and decision-making~\cite{schwarzschild2021}. Adversaries can manipulate Deep Neural Networks (DNNs) to serve malicious purposes by corrupting training data. Research indicates that poisoned data can be discreetly added to datasets during training or fine-tuning of Language Models (LMs), exploiting vulnerabilities in data pipelines~\cite{kurita2020}. These risks are particularly pronounced when using external or unverified dataset sources, highlighting the need for strict data curation and validation to mitigate such threats.

\BfPara{Defense Techniques} 
Several defense techniques have been employed to protect LLMs from poisoning attacks, including data validation, filtering, cleaning, and anomaly detection~\cite{Shah2022}. Yan~\etal~\cite{yan2023} introduced ParaFuzz, a framework designed to detect poisoned samples in NLP models. This technique employs fuzzing, a software testing methodology, to identify poisoned samples as outliers by analyzing the interpretability of model predictions. ParaFuzz provides a robust method for filtering malicious data during training, noting that poisoned data often diverges from benign distributions and takes longer for a model to learn. Additionally, dataset curation techniques, as highlighted by Contiella~\etal~\cite{Continella2017}, emphasize the importance of removing near-duplicate poisoned samples, identifying known triggers, and isolating anomalies in training datasets. This approach has proven effective against attacks such as AutoPoison~\cite{shu2023} and TrojanPuzzle~\cite{aghakhani2024}. Fine-pruning has also emerged as an effective defense against data poisoning attacks~\cite{liu2018}, while perplexity filtering and query rephrasing are specifically employed to mitigate white-box attacks such as AgentPoison~\cite{chen2024a}. Despite these advancements, continuous research is necessary to refine and optimize defenses against evolving threats in modern LLMs.

\subsubsection{Backdoor Attacks} 
In a backdoor attack, poisoned samples are introduced into a model to embed hidden malicious functionality. These attacks allow the model to perform normally on benign inputs but behave maliciously on specific, poisoned inputs. Backdoor attacks in LLMs can be categorized as input-triggered, prompt-triggered, instruction-triggered, and demonstration-triggered~\cite{chen2024b}. Adversaries use techniques such as injecting triggers into training data, modifying prompts to elicit malicious output, exploiting fine-tuning processes with poisoned instructions, or subtly altering demonstrations to manipulate model behavior while embedding hidden vulnerabilities without detection~\cite{chowdhury2024}.

\BfPara{Defense Techniques} 
To counter backdoor attacks in LMs, various advanced mitigation techniques have been proposed. One such approach is \textit{Fine Mixing}, introduced by Zhang~\etal~\cite{zhang2022}, which employs a two-step fine-tuning process that merges backdoor-optimized weights with pre-trained weights, followed by refinement on a clean dataset. This method also integrates \textit{Embedding Purification} (E-PUR) to neutralize backdoors in word embeddings, improving model robustness. Another technique, \textit{CUBE} (Clustering-Based Unsupervised Backdoor Elimination) by Cui~\etal~\cite{cui2022}, leverages the HDBSCAN density clustering algorithm to identify and separate poisoned samples from clean ones based on distinct clustering patterns. Additionally, \textit{Masking Differential Prompting} (MDP) by Xi~\etal~\cite{xi2023} offers an efficient and adaptable defense for prompt-based LMs by exploiting the increased sensitivity of poisoned samples to random masking, which causes significant variations in their probability distributions. While these techniques enhance security, further research is needed to assess their effectiveness against advanced backdoor attacks in LLMs such as GPT-4 and Llama-3.

\subsection{Prompt Hacking} 
Prompt hacking involves manipulating input prompts to influence the output of LLMs. By crafting precise and intentional prompts, attackers aim to direct model responses toward specific objectives, which may include generating unintended or harmful outcomes. Since LLMs operate through interaction-based systems where user queries drive their outputs, carefully designed prompts can exploit the model's underlying mechanisms, overriding safeguards and producing misleading, malicious, or unexpected results. 

\subsubsection{Jailbreaking Attacks} 
Jailbreaking attacks involve bypassing software restrictions imposed by manufacturers or service providers, granting users elevated access to system functionality. While commonly associated with Apple's iOS~\cite{wolk2009}, similar practices exist for Android and other systems. Jailbreaking grants users privileged access to core functions and the file system, allowing for unauthorized application installations, bypassing regional locks, and performing advanced system manipulations~\cite{mondillo2025}. However, it introduces significant risks, such as compromised security, loss of functionality, and potentially irreversible damage to the device.

\BfPara{Defense Techniques} 
Several defense techniques have been developed to mitigate jailbreaking attacks on LLMs. Kumar~\etal~\cite{kumar2024} introduced \textit{substring safety filtering}, which analyzes and filters input prompts to block harmful or unintended responses, providing robust defense despite its increased complexity for longer inputs. Wu~\etal~\cite{wu2023} developed a \textit{self-reminder system} that directs LLMs toward safe behaviors, improving context-specific responses and reducing jailbreak success rates, particularly in role-playing scenarios. Jin~\etal~\cite{JinHLZCZ24} proposed a \textit{goal prioritization} method that prioritizes safety over utility in response generation, thereby reducing harmful content risks. Additionally, Robey~\etal~\cite{robey2024} introduced the \textit{Smooth LLM} framework, which applies randomized smoothing by perturbing input prompts and aggregating outputs to lower the success rate of instruction-based attacks on models such as Llama-2 and Vicuna, enhancing model defenses against emerging threats.

\subsubsection{Prompt Injection} 
Prompt injection manipulates LLMs to generate attacker-desired outputs by bypassing safety mechanisms~\cite{crothers2023}. Carefully crafted prompts enable adversaries to override original commands or execute malicious actions. This vulnerability facilitates harmful content generation, including data leakage, unauthorized access, hate speech, disinformation, and other security breaches~\cite{rababah2024}. In prompt injection attacks, adversaries may directly instruct the LLM to bypass filtering mechanisms or process compromised inputs. Additionally, attackers can pre-inject harmful prompts into web content, which the LLM may inadvertently process, making these attacks difficult to detect and mitigate.

\BfPara{Defense Techniques} 
Prompt injection defenses are categorized into prevention-based and detection-based methods, with ongoing research efforts enhancing their effectiveness. Prevention-based defenses, as described in~\cite{learnprompting2023}, seek to block injected tasks through techniques such as \textit{paraphrasing}, \textit{re-tokenization}~\cite{jain2023}, and \textit{data prompt isolation}~\cite{schulhoff2024}. Paraphrasing disrupts the sequence of injected data, while re-tokenization breaks down infrequent tokens in compromised prompts, mitigating malicious instructions. Detection-based defenses, such as those proposed by Wang~\etal~\cite{Wang2023}, assess prompt integrity through response-based or prompt-based evaluations. \textit{Perplexity-based detection}~\cite{gonen2024}, for instance, identifies compromised triggers by analyzing quality degradation and increased perplexity. Despite these advancements, Liu~\etal~\cite{liu2024d} found that traditional prevention and detection techniques remain inadequate against optimization-based attacks, such as \textit{Judge Deceiver}~\cite{shi2024}, highlighting the need for continuous innovation in this field.

\begin{table*}[t]
\centering    
\caption{
Comprehensive Analysis of Security Defense Techniques and Related Approaches. 
\textbf{Abbreviations:} 
RT (Red Team), 
CF (Content Filtering), 
SFT (Safty Fine Tuning), 
MM (Model Merge), 
CE (CUBE), 
GP (Goal Prioritization), 
FM (Fine Mixing), 
P (Perplexity), 
PF (ParaFuzz), 
SF (Substring Filtering), 
S (Smooth), 
DPI (Data Prompt Isolation), 
PH (Paraphrasing), 
SR (Self Reminder), 
C (Cleaning), 
CU (Curation), 
MDP (Masking Differential Prompting), 
R (Re-tokenization),
PI (Prompt Injection).}
\label{table:techniques}
\vspace{-2mm}
\begin{threeparttable}
    \centering
    \scalebox{0.95}{\begin{tabular}{llp{0.4mm}*{19}{p{2.5mm}}}
    \Xhline{2\arrayrulewidth}
    \textbf{Author} & \textbf{Ref.}&
    \rotatebox[origin=l]{90}{RT} &  
    \rotatebox[origin=l]{90}{CF} &  
    \rotatebox[origin=l]{90}{SFT} &  
    \rotatebox[origin=l]{90}{MM} &  
    \rotatebox[origin=l]{90}{CE} &  
    \rotatebox[origin=l]{90}{GP} &  
     \rotatebox[origin=l]{90}{FM} &  
    \rotatebox[origin=l]{90}{P} &  
    \rotatebox[origin=l]{90}{PF} &  
    \rotatebox[origin=l]{90}{SF} &  
    \rotatebox[origin=l]{90}{S} &  
    \rotatebox[origin=l]{90}{DPI} &  
    \rotatebox[origin=l]{90}{PH} &  
    \rotatebox[origin=l]{90}{SR} &  
    \rotatebox[origin=l]{90}{C} &  
    \rotatebox[origin=l]{90}{CU} &  
    \rotatebox[origin=l]{90}{MDP} &  
    \rotatebox[origin=l]{90}{R} &  
     \rotatebox[origin=l]{90}{PI}\\ 
    
    \Xhline{1\arrayrulewidth}
     Alone~\etal&\cite{alon2023} &\ding{53} &\cmark &\ding{53} &\ding{53}&\ding{53} &\ding{53}&\ding{53} &\cmark &\ding{53}&\ding{53} &\ding{53} &\ding{53}&\ding{53} &\ding{53} &\ding{53} &\ding{53} &\ding{53} &\ding{53} &\ding{53} \\
      Bianchi~\etal &\cite{BianchiSA24}&\ding{53}&\ding{53}&\cmark &\ding{53} &\ding{53} &\ding{53} &\ding{53}&\ding{53} &\ding{53}&\cmark &\ding{53} &\ding{53} &\ding{53}&\ding{53} &\ding{53} &\ding{53} &\ding{53} &\ding{53}&\ding{53} \\
       Cao~\etal &\cite{CaoXHZ2022}  &\ding{53} &\cmark &\ding{53} &\ding{53} &\ding{53} &\ding{53}&\ding{53} &\cmark  &\ding{53}&\ding{53} &\ding{53} &\ding{53}&\ding{53} &\ding{53} &\ding{53} &\ding{53} &\ding{53} &\ding{53}&\ding{53} \\
        Contiella~\etal &\cite{Continella2017}&\ding{53} &\ding{53} &\ding{53} &\ding{53}&\ding{53} &\ding{53}&\ding{53} &\ding{53} &\ding{53}&\ding{53} &\ding{53} &\ding{53}&\ding{53} &\ding{53} &\ding{53} &\cmark &\ding{53} &\ding{53}&\ding{53} \\
         Cui~\etal&\cite{cui2022}&\ding{53} &\ding{53} &\ding{53} &\ding{53}&\cmark  &\ding{53}&\ding{53} &\ding{53} &\ding{53}&\ding{53} &\ding{53} &\ding{53}&\ding{53} &\ding{53} &\ding{53} &\ding{53} &\ding{53} &\ding{53}&\ding{53} \\
         Ribeiroet~\etal &\cite{ganguli2022} &\cmark &\ding{53} &\ding{53} &\ding{53}&\ding{53} &\ding{53}&\ding{53} &\ding{53} &\ding{53}&\ding{53} &\ding{53} &\ding{53}&\ding{53} &\ding{53} &\ding{53} &\ding{53} &\ding{53} &\ding{53} &\ding{53}\\
          Ganguali~\etal &\cite{GanguliLKABK22} &\cmark &\ding{53} &\ding{53} &\ding{53}&\ding{53} &\ding{53}&\ding{53} &\ding{53} &\ding{53}&\ding{53} &\ding{53} &\ding{53}&\ding{53} &\ding{53} &\ding{53} &\ding{53} &\ding{53} &\ding{53} &\ding{53}\\
          Gonen~\etal&\cite{gonen2024} &\ding{53} &\ding{53} &\ding{53} &\ding{53}&\ding{53} &\ding{53}&\ding{53} &\cmark  &\ding{53}&\ding{53} &\ding{53} &\ding{53}&\ding{53} &\ding{53} &\ding{53} &\ding{53} &\ding{53} &\ding{53} &\ding{53}\\
           Jain~\etal&\cite{jain2023} &\ding{53} &\cmark &\ding{53} &\ding{53} &\ding{53} &\ding{53}&\ding{53} &\ding{53} &\ding{53}&\ding{53} &\ding{53} &\ding{53}&\cmark &\ding{53}&\ding{53}&\ding{53} &\ding{53} &\cmark &\ding{53} \\
            Jin~\etal&\cite{JinHLZCZ24} &\ding{53} &\ding{53} &\ding{53} &\ding{53}&\ding{53}&\cmark &\ding{53}  &\ding{53} &\ding{53}&\ding{53} &\ding{53} &\ding{53}&\ding{53} &\ding{53} &\ding{53} &\ding{53} &\ding{53}&\ding{53}&\ding{53}\\
             Kupmar~\etal&\cite{kumar2024}&\ding{53} &\cmark  &\ding{53} &\ding{53}&\ding{53} &\ding{53}&\ding{53} &\ding{53} &\ding{53}&\cmark &\ding{53} &\ding{53}&\ding{53} &\ding{53} &\ding{53} &\ding{53} &\ding{53} &\ding{53} &\ding{53}\\
              Liu~\etal&\cite{liu2024a}&\ding{53} &\ding{53}&\ding{53}&\ding{53} &\ding{53}&\ding{53} &\ding{53}&\cmark&\ding{53} &\ding{53} &\ding{53} &\ding{53}& \cmark &\ding{53} &\ding{53} &\ding{53} &\ding{53}&\ding{53} &\ding{53} \\
 Perez~\etal&\cite{perez2022} &\cmark &\ding{53} &\ding{53} &\ding{53}&\ding{53} &\ding{53}&\ding{53} &\ding{53} &\ding{53}&\ding{53} &\ding{53} &\ding{53}&\ding{53} &\ding{53} &\ding{53} &\ding{53} &\ding{53} &\ding{53} &\ding{53}\\
  Xiangyu~\etal &\cite{qi2024}  &\ding{53}&\ding{53}&\cmark &\ding{53} &\ding{53} &\ding{53} &\ding{53}&\ding{53} &\ding{53}&\cmark &\ding{53} &\ding{53} &\ding{53}&\ding{53} &\ding{53} &\ding{53} &\ding{53} &\ding{53}&\ding{53} \\
   Robey~\etal&\cite{robey2024} &\ding{53} &\ding{53} &\ding{53} &\ding{53}&\ding{53} &\ding{53}&\ding{53} &\ding{53} &\ding{53}&\ding{53} &\cmark &\ding{53}&\ding{53} &\ding{53} &\ding{53} &\ding{53} &\ding{53}&\ding{53}&\ding{53} \\
    Schulhoff~\etal&\cite{schulhoff2024} &\ding{53} &\ding{53} &\ding{53} &\ding{53}&\ding{53} &\ding{53}&\ding{53} &\ding{53} &\ding{53}&\ding{53} &\ding{53} &\cmark&\cmark &\ding{53} &\ding{53} &\ding{53} &\ding{53} &\ding{53}&\ding{53} \\
    Shan~\etal &\cite{Shah2022} & \ding{53} &\ding{53} &\ding{53} &\ding{53} &\ding{53}&\cmark &\ding{53} &\ding{53} &\ding{53} &\ding{53} &\ding{53}&\ding{53} &\ding{53} &\ding{53}&\ding{53} &\cmark &\ding{53} &\ding{53} &\ding{53} \\
   Wang~\etal&\cite{Wang2023} &\ding{53} &\ding{53} &\ding{53} &\ding{53}&\ding{53} &\ding{53}&\ding{53} &\ding{53} &\ding{53}&\ding{53} &\ding{53} &\ding{53}&\ding{53}&\ding{53} &\ding{53} &\ding{53}  &\ding{53} &\ding{53} &\cmark \\
   Wu~\etal &\cite{wu2023} &\ding{53} &\ding{53} &\ding{53} &\ding{53}&\ding{53} &\ding{53}&\ding{53} &\ding{53} &\ding{53}&\ding{53} &\ding{53} &\ding{53}&\ding{53} &\cmark &\ding{53} &\ding{53} &\ding{53} &\ding{53} &\ding{53}\\
   Xi~\etal&\cite{xi2023}&\ding{53} &\ding{53} &\ding{53} &\ding{53}&\ding{53} &\ding{53}&\ding{53} &\ding{53} &\ding{53}&\ding{53} &\ding{53} &\ding{53}&\ding{53} &\ding{53} &\ding{53} &\ding{53} &\cmark &\ding{53}&\ding{53}\\
    Yan~\etal&\cite{yan2023} &\ding{53} &\ding{53} &\ding{53} &\ding{53}&\ding{53} &\ding{53}&\ding{53} &\ding{53} &\cmark&\cmark &\ding{53} &\ding{53}&\ding{53} &\ding{53} &\ding{53} &\ding{53} &\ding{53}&\ding{53}&\ding{53} \\
      Zhang~\etal&\cite{zhang2022} &\ding{53} &\ding{53} &\ding{53} &\ding{53}&\ding{53} &\ding{53}&\cmark &\ding{53} &\ding{53}&\ding{53} &\ding{53} &\ding{53}&\ding{53} &\ding{53} &\cmark &\ding{53} &\ding{53} &\ding{53} &\ding{53}\\
       Zhao~\etal&\cite{ZhaoDMZR24} &\ding{53} &\ding{53}&\cmark &\ding{53} &\ding{53} &\ding{53} &\ding{53}&\ding{53} &\ding{53}&\ding{53} &\ding{53} &\ding{53} &\ding{53}&\ding{53} &\ding{53} &\ding{53} &\ding{53} &\ding{53} &\ding{53}\\
          
         Zou~\etal&\cite{zou2023} &\ding{53} &\ding{53} &\ding{53} &\cmark&\ding{53} &\ding{53}&\ding{53} &\ding{53} &\ding{53}&\ding{53} &\cmark &\ding{53}&\ding{53} &\ding{53} &\ding{53} &\ding{53} &\ding{53} &\ding{53}&\ding{53} \\
         \Xhline{2\arrayrulewidth}

        \end{tabular}}
\end{threeparttable}
\end{table*}

\autoref{table:techniques} presents an analysis of security techniques employed across various approaches to safeguard LLMs from security attacks. This analysis highlights the diverse methodologies adopted in existing research and provides insights into their effectiveness and robustness in protecting LLMs.

\section{Limitations and Future Directions}\label{sec:Limitations}

While LLMs have shown significant potential in addressing cybersecurity challenges, several inherent limitations hinder their broader adoption and effectiveness in security tasks. One major limitation is the lack of interpretability, as the black-box nature of LLMs prevents users from understanding how models make security-critical decisions. This undermines trust and transparency, which are essential for deployment in sensitive domains. The lack of transparency is particularly concerning due to the increased risks posed by AI-generated content (AIGC)~\cite{raiaan2024}, including privacy breaches, the spread of misinformation, and the production of vulnerable code~\cite{mishra2024}. Moreover,  LLMs's use in cybersecurity domains such as network, hardware, blockchain, and content security is constrained by the lack of high-quality, domain-specific datasets needed for effective fine-tuning~\cite{zhao2024}. Future research should focus on developing explainability tools to clarify model decisions, collaborating with domain experts to curate relevant datasets, and refining models to integrate cybersecurity-specific knowledge. Expanding LLM capabilities to handle multimodal inputs such as voice, images, and videos would enhance their contextual understanding in security settings~\cite{xu2024}. Addressing these challenges will enable LLMs to better support cybersecurity efforts, offering deeper insights and facilitating the development of secure, automated solutions.

Despite substantial advancements in understanding security vulnerabilities in machine learning models, research on mitigating backdoor attacks in LLM-based tasks, such as text summarization and generation, remains limited~\cite{yang2024}. Mitigating backdoor attacks is crucial for ensuring robust defenses and secure LLM deployment. While poisoning attacks on ML models have been well studied~\cite{moore2023}, advanced techniques such as ProAttack~\cite{Zhao2023} and BadPrompt~\cite{cai2022} are not yet fully addressed.

To enhance LLM security and build robust systems, further research is required across various tasks and architectures. Current defense measures, such as dataset cleansing~\cite{huang2022} and anomaly detection, often disrupt the development process by inadvertently removing critical data. Similarly, methods like early stopping after specific training epochs offer moderate protection but frequently lead to reduced model performance~\cite{wan2023}. Adaptive defense mechanisms are necessary to strike a balance between securing models and maintaining their utility.

Despite their potential, LLMs remain inherently vulnerable to security risks, posing critical challenges for their application in cybersecurity tasks. One key avenue for future research is enabling LLMs to autonomously detect and resolve vulnerabilities within their architectures. Such self-repair capabilities would enhance resilience while reducing dependence on external restrictions. Achieving this requires a dual focus: automating cybersecurity tasks using LLMs while concurrently implementing robust self-protection mechanisms to mitigate model-specific risks. Current research highlights significant shortcomings in LLM defenses. For example, while safety features in existing models, such as those in ChatGPT, can prevent simple attacks, multi-step exploits continue to compromise these systems~\cite{Martin23}. Moreover, newer AI systems, such as Bing AI, have demonstrated even greater susceptibility to advanced attack strategies~\cite{wang2019}. Techniques inspired by human reasoning, such as self-reminder mechanisms~\cite{das2024}, have been proposed to mitigate these vulnerabilities; however, their effectiveness in handling complex queries remains underexplored. Addressing this gap requires deeper insights into how these mechanisms influence model reasoning and further refinements to optimize defense strategies without compromising performance.

Defending LLMs against vulnerabilities requires addressing both intrinsic weaknesses and external adversarial threats. Given the impracticality of manually auditing training data sets, alternative strategies have been used, such as personal information filtering and restrictive terms of use, to mitigate the inclusion of sensitive content. Advanced techniques like MDP~\cite{xi2023} have shown promise in countering earlier backdoor attacks. However, these methods struggle with complex NLP tasks, such as paraphrasing and semantic similarity, and do not fully address emerging threats like BadPrompt~\cite{cai2022} and BToP~\cite{yao2024}. The lack of comprehensive evaluation metrics, such as perplexity, further complicates the assessment of attack and defense strategies. Additionally, the high computational demands of LLM training in federated learning (FL) environments create additional obstacles to scalable and adaptive defenses. 

Organizations such as OpenAI implement defense measures, yet these do not fully eliminate risks like jailbreaking. Malicious datasets and prompts still allow attackers to bypass security mechanisms and generate harmful content~\cite{charfeddine2024}. These persistent vulnerabilities underscore the urgent need for flexible and scalable defense strategies that can adapt to evolving attack techniques.

In summary, securing LLMs requires a holistic strategy to improve robustness, scalability, and effectiveness across various applications. Future studies should focus on developing interpretability frameworks, autonomous protection mechanisms, multimodal capabilities, and improved evaluation metrics to create more secure and reliable AI-driven cybersecurity solutions.

\section{Conclusion} \label{sec:Conclusion}

LLMs are at the forefront of transforming cybersecurity, offering innovative solutions to address increasingly complex challenges. This survey analyzed LLM applications on 32 security tasks covering eight domains, including blockchain, hardware, IoT, and cloud security, with a focus on task-specific applications such as vulnerability detection, malware analysis, and threat intelligence automation. Their versatility and significant impact were demonstrated in various cybersecurity applications.

This study initially explored the landscape of LLM applications, categorizing security tasks within each domain while highlighting their potential in modern cybersecurity. Furthermore, we examined LLM vulnerabilities, such as adversarial attacks and prompt injection, and identified mitigation strategies, including re-tokenization, perplexity-based detection, prompt isolation, and other defense mechanisms aimed at enhancing security.

This survey lays the foundation for future research, providing key insights for integrating LLMs into secure cybersecurity frameworks. By addressing existing challenges, LLMs can evolve to develop robust solutions that counter emerging cyber threats and safeguard critical digital infrastructure.

\end{document}